# Robust and clean Majorana zero mode in the vortex core of high-temperature superconductor (Li$_{0.84}$Fe$_{0.16}$)OHFeSe

Qin Liu[1,2,3+], Chen Chen[1,3+], Tong Zhang[1,3*], Rui Peng[1,3], Ya-Jun Yan[1,3], Chen-Hao-Ping Wen[1,3], Xia Lou[1,3], Yu-Long Huang[4], Jin-Peng Tian[4], Xiao-Li Dong[4], Guang-Wei Wang[5], Wei-Cheng Bao[6,7], Qiang-Hua Wang[3,6], Zhi-Ping Yin[5*], Zhong-Xian Zhao[4], Dong-Lai Feng[1,3*]

[1] State Key Laboratory of Surface Physics, Department of Physics, and Advanced Materials Laboratory, Fudan University, Shanghai 200438, China
[2] Science and Technology on Surface Physics and Chemistry Laboratory, Mianyang, Sichuan 621908, China
[3] Collaborative Innovation Center of Advanced Microstructures, Nanjing 210093, China
[4] Beijing National Laboratory for Condensed Matter Physics and Institute of Physics, Chinese Academy of Sciences, Beijing 100190, China
[5] Department of Physics and Center for Advanced Quantum Studies, Beijing Normal University, Beijing 100875, China
[6] National Laboratory of Solid State Microstructures & School of Physics, Nanjing University, Nanjing, 210093, China
[7] Zhejiang University of Water Resources and Electric Power, Hangzhou 310018, China

[+] These authors contributed equally.
*Email: tzhang18@fudan.edu.cn, yinzhiping@bnu.edu.cn, dlfeng@fudan.edu.cn

The Majorana fermion, which is its own anti-particle and obeys non-abelian statistics, plays a critical role in topological quantum computing. It can be realized as a bound state at zero energy, called a Majorana zero mode (MZM), in the vortex core of a topological superconductor, or at the ends of a nanowire when both superconductivity and strong spin orbital coupling are present. A MZM can be detected as a zero-bias conductance peak (ZBCP) in tunneling spectroscopy. However, in practice, clean and robust MZMs have not been realized in the vortices of a superconductor, due to contamination from impurity states or other closely-packed Caroli-de Gennes-Matricon (CdGM) states, which hampers further manipulations of MZMs. Here using scanning tunneling spectroscopy, we show that a ZBCP well separated from the other discrete CdGM states exists ubiquitously in the cores of free vortices in the defect free regions of (Li$_{0.84}$Fe$_{0.16}$)OHFeSe, which has a superconducting transition temperature of 42 K. Moreover, a Dirac-cone-type surface state is observed by angle-resolved photoemission spectroscopy, and its topological nature is confirmed by band calculations. The observed ZBCP can be naturally attributed to a MZM arising from this chiral topological surface states of a bulk superconductor. (Li$_{0.84}$Fe$_{0.16}$)OHFeSe thus provides an ideal platform for studying MZMs and topological quantum computing.

## I. INTRODUCTION

In recent years, many recipes have been proposed for realizing MZMs [1-3], as a critical step towards topological quantum computation [4,5]. For example, MZMs are predicted to exist at the ends of a semiconductor nanowire with strong spin-orbital coupling (SOC), when it is in proximity to a superconductor and under a sufficiently large Zeeman field [2-3,6-8]. A quantized ZBCP has been observed in the tunneling spectrum of a hybrid device between superconducting aluminum and an InSb nanowire [9], providing compelling evidence for a MZM. However, fundamental quantum computing operations such as braiding of MZMs in nanowires [10] are challenging and yet to be realized after a tremendous amount of ingenious technical endeavor [11,12].

Topological systems with additional dimensions provide a broader hunting ground in the search of MZM, e.g. in the vortex cores of a topological superconductor or superconducting heterostructures [1-3]. For conventional $s$-wave superconductors, confined quasi-particles in the vortex core give rise to CdGM bound states with $E=\mu\Delta^2/E_F$, where $\mu$ is a half integer ($\pm 1/2$, $\pm 3/2$...) [13]. However, for a chiral $p$-wave superconductor, bound states still exist but with integer $\mu$ (0, $\pm 1$, $\pm 2$...) – the $E=0$ state, or zero-energy mode, is a MZM [14]. Chiral $p$-wave superconductors are extremely scarce, but Fu and Kane have proposed that proximity effects from an $s$-wave superconductor on topological surface states would produce a two-dimensional system whose Hamiltonian effectively resembles a spinless $p\pm ip$ superconductor, and thus it can host Majorana bound states in its vortices [15] (The spinless nature is due to the fact that topological surface state is spin non-degenerate). Based on this scenario, ZBCPs that may potentially correspond to MZMs have been found in the vortex cores of topological insulator / superconductor heterostructures (e.g. $Bi_2Te_3/NbSe_2$ and $Bi_2Te_3/FeTe_xSe_{1-x}$) [16-18], and topological surface states of bulk superconductors (e.g. $Cu_xBi_2Se_3$ and $FeTe_xSe_{1-x}$) [19-22]. However, an unambiguous identification or isolation of a MZM is still lacking in these systems. For materials with a large Fermi energy ($E_F$), such as $Bi_2Se_3/NbSe_2$, $Cu_xBi_2Se_3$ and $Bi_2Te_3/FeTe_xSe_{1-x}$, a mode at zero energy is buried under many densely packed CdGM states in the vortex core, their energies separated by only a few $\mu$eV. $FeTe_xSe_{1-x}$ has a small $E_F$, so its ZBCP can be readily distinguished from the CdGM states. However, there are intrinsic impurity effects of interstitial Fe and heavy Te doping [23,24]. A zero-bias impurity state was observed on interstitial Fe which is insensitive to magnetic field [23], but whether this is related to a MZM is yet to be confirmed. A ZBCP can be observed in a fraction of vortices in certain $FeTe_xSe_{1-x}$ samples [21], while it cannot be observed in samples annealed to reduce impurities [24]. Therefore, to study the largely-unknown properties of MZMs, one needs to find a system that can host MZMs in a clean and robust manner, with weak pinning and scattering from impurities, which would lay the foundation for further operations such as braiding of MZMs.

To reach these goals, we need a superconducting system that has a large superconducting gap, a small $E_F$, a stoichiometric lattice (at least in the sublattice responsible for superconductivity) with few defects, a short superconducting coherence length ($\xi$) to potentially reduce vortex pinning, and a topologically nontrivial electronic structure. The large gap usually also means high superconducting transition temperature ($T_c$) and large separations between bound states in vortices, which will make the operation temperature easier to reach in practice. As topological non-trivial band structures have been predicted to exist in many iron-based superconductors [25-27], $(Li_{1-x}Fe_x)OHFeSe$ with a $T_c$ as high as 42 K appears to be a promising candidate. It belongs to the heavily-electron-doped family of iron selenide that only have electron Fermi pockets, and its $E_F$ is merely 50~60 meV [28,29]. Unlike FeTe$_x$Se$_{1-x}$, the superconducting FeSe layers in (Li$_{1-x}$Fe$_x$)OHFeSe are stoichiometric (Fig. 1(a)). The remaining questions are whether it could exhibit topologically nontrivial behavior and if it hosts MZMs. To address these questions, we conduct density functional theory (DFT) and dynamical mean field theory (DMFT) calculations and ARPES measurements on (Li$_{0.84}$Fe$_{0.16}$)OHFeSe, and we find non-trivial band inversion and topological surface states. We studied the vortex states of (Li$_{0.84}$Fe$_{0.16}$)OHFeSe by low-temperature scanning tunneling microscopy (STM). In the core of impurity-free vortices, we indeed discovered discrete CdGM states which are clearly separated from a unique state located exactly at zero bias. Such a *clean and robust* zero energy mode exists in all the free vortex cores of (Li$_{0.84}$Fe$_{0.16}$)OHFeSe and matches all expectations of Fu and Kane's theory, providing an ideal platform to further explore the properties and applications of MZMs.

## II. RESULTS

### A. Electronic structure calculations and ARPES measurements

In Fig. 1(b), we present our DFT combined with DMFT calculations of the electronic structure of (Li$_{0.75}$Fe$_{0.25}$)OHFeSe (details of the calculation are described in Appendix and section **S1** of Supplementary Materials). Along the $\Gamma$ - $Z$ direction, the three flat bands around the Fermi level are dominated by the Fe *3d$_{xy}$*, *3d$_{xz}$* and *3d$_{yz}$* orbitals whereas the dispersive band is composed of Se *4p$_z$* orbital and the *3d$_z^2$* orbital of Fe in the (Li$_{1-x}$Fe$_x$)OH layer. This dispersive Se *4p$_z$* band has odd parity at the $\Gamma$ and $Z$ points. It crosses the Fermi level and the Fe *3d* bands along $\Gamma - Z$, giving rise to a non-trivial band inversion and band topology. Here Fe atoms in the (Li$_{1-x}$Fe$_x$)OH layer play an important role as they hybridize strongly with the Se *4p$_z$* orbital and change the dispersion and position of the Se *4p$_z$* band. Without the Fe atom in the (Li$_{1-x}$Fe$_x$)OH layer, the band inversion along the $\Gamma$ - $Z$ direction would not have existed and the band topology would have been trivial (as demonstrated in Fig. S1(b)). The spin-orbital coupling (SOC) also plays an important role. Without SOC, the Fe *3d$_{xz}$* and *3d$_{yz}$* bands are degenerate at $\Gamma$ point and are in the $\Gamma_5^+$ states, whereas the dispersive Se *4p$_z$* band is in the odd $\Gamma_2^-$ state at $\Gamma$ point. With SOC, the doubly degenerate $\Gamma_5^+$ states are separated to a singlet low-lying $\Gamma_6^+$ state and an upper $\Gamma_7^+$ state, and the $\Gamma_2^-$ state changes to $\Gamma_6^-$ state [26]. Under the C$_{4v}$ symmetry inherent in the tetragonal crystal structure, the two $\Gamma_6^+$ and $\Gamma_6^-$ state-derived $\Lambda_6$ bands with dominating Fe *3d$_{xz}$* and Se *4p$_z$* orbital characters hybridize with each other along $\Gamma$ - $Z$ direction, and open a ~2.5 meV gap around their crossing point (marked by the circle in Fig. 1(b)). The $Z_2$ invariant is calculated to be 1 after this SOC gap opens [26], which suggests (Li$_{0.75}$Fe$_{0.25}$)OHFeSe is in a topologically non-trivial phase. Since one cannot make a transition between two states with different topologies without closing the band gap, namely, here between a topological nontrivial bulk with band inversion and the topological trivial "insulating" vacuum, the transition region, i.e. surface, would hosts gapless topological surface states (TSS) in the band gap. More specifically, the hybridization gap between the $\Gamma_6^+$ and $\Gamma_6^-$ states at the surface are removed by the discontinuity of the surface. Indeed, our calculations show that Dirac-cone-like surface states centered at $\bar{\Gamma}$ *appear* on the (001) surface as shown in Fig. 1(c). The topological surface states have helical spin texture and can produce the MZM when it is in close proximity to s-wave superconductivity [15, 26, 27].

Guided by these calculations, we have conducted further ARPES measurements on high quality (Li$_{0.84}$Fe$_{0.16}$)OHFeSe films grown on LaAlO$_3$ by hydrothermal epitaxy ($T_c$~42 K) [30,31] (see Appendix for details). Fig. 1(d) shows the ARPES spectrum of a cleaved film across the Brillouin zone (BZ) center as shown in Fig. 1(a). Multiple parabolic bands can be clearly resolved below the Fermi surface, and there is one flat band around -300 meV. Moreover, an electron band with its band bottom at ~ -50meV is observed at $M$ (Fig. 1(j, k)). These observed bands and the calculated ones in Fig. 1(b) have one-to-one correspondence, although qualitatively. The relative positions and bandwidths may differ due to correlation effects. We also note that certain bands, such as the other electron-like band at $M$, were not observed here due to known matrix element effect. A precise band calculation for heavily electron doped iron selenides is still so far challenging, however, qualitative features of the measured bands are reproduced by the calculation in Fig. 1(b), and the generic features here resemble several iron chalcogenide superconductors [32].

Intriguingly, in the region near the Fermi energy as plotted in Fig. 1(e), some finite spectral weight within the bulk band gap can be clearly observed. The second derivative of the region containing these spectral weight (Fig. 1(f)) display an "X"-shaped band structure. Fig. 1(g) shows the momentum distribution curves (MDCs) of the data in Fig. 1(e) near $E_F$. Through one-Lorentzian-peak fit (shown in Fig. S2(a)), we found the width of the spectral peak near $k_{//}=0$ has a minimum at $E=-20$ meV (Fig. 1(h)). This would indicates a position of band crossing point. Then we apply two-peak fit to MDCs with assuming each peak has a constant FWHM as that of the single peak at $E=-20$ meV (the scattering rate variations are expected to be negligible in such a small energy window). The results display a nearly-linear, Dirac-cone like band dispersion, as shown in Fig. 1(g), and its overall fitting quality is better than that of the one-peak fit (see Fig. S2(c) for comparison). In Fig. 1(i) we plot the E-k relations extracted from the two-peak fit, a linear fit is found to be plausible near the crossing point which gives a $v_F = 5.5(\pm1.5)\times10^4$ m/s and $E_D= 20(\pm2)$ meV (see Fig. S2(d) for more details). The $k_F$ is determined to be ~0.028 Å$^{-1}$ through the MDC near $E_F$.

We note the existence of topological surface states is determined by the band topology. The qualitative agreement between the measured and calculated band structures throughout BZ would suggest they have the same topology. Therefore, the predicted topological surface states provide the most natural explanation for such a nearly-linear, Dirac-cone-like band dispersion at $\Gamma$. Nevertheless, one should examine whether it is related to any other known electron pockets in the (Li$_{1-x}$Fe$_x$)OH and FeSe surface layers exposed by cleavage. On the FeSe surface layer, an electron-like pocket would only appear around $\Gamma$ after heavily doping electrons through potassium dosing, and thus it is well above the Fermi energy in the case here [33]. There is possibly an electron pocket around $\Gamma$ in the (Li$_{1-x}$Fe$_x$)OH surface layer, based on our previous quasi-particle interference (QPI) measurements of (Li$_{0.8}$Fe$_{0.2}$)OHFeSe bulk crystal [34]. However, its band bottom is at about 50 meV below $E_F$ with a much larger Fermi velocity, so it cannot be the Dirac cone observed here. Therefore, we conclude that the observed Dirac band dispersion can only be attributed to the topological surface state predicted in our calculations.

The FeSe surface of (Li$_{1-x}$Fe$_x$)OHFeSe is fully gapped [34,35], as shown in Fig. 2 below. If this topological surface state is on the FeSe surface, it thus should be gapped. A superconducting gap of about 10 meV can be observed on the electron-like band around $\bar{M}$, as shown in Fig. 1(j, k), and further illustrated in Fig. 1(l) (details described in caption). The absence of gap in our data for the topological surface state around $\bar{\Gamma}$ may be attributed to the fact that this gap is small, while the resolution of ARPES is about 6 meV. Moreover, ARPES collects data over a millimeter-sized region and thus contains large contribution from the non-superconducting and disordered $(Li_{1-x}Fe_x)OH$ surface, which has low energy spectral weight near $\bar{\Gamma}$.

### B. STM characterization of the sample surface and superconducting gap

STM measurement was conducted in a cryogenic STM with a base temperature of 0.4 K and had an energy resolution of 0.36 meV (see Appendix and Supplementary Materials section **S3**). Figure 2(a) shows typical topography of a cleaved $(Li_{0.84}Fe_{0.16})OHFeSe$ film. We observed both FeSe-terminated (left) and $(Li_{1-x}Fe_x)OH$-terminated (right) surfaces. The Fig. 2(a) inset shows the atomically-resolved FeSe lattice with $a_0$ = 3.8 Å. Dimer-like defects at the Fe site were frequently observed (dashed circle), which are likely substitutional impurities. The typical tunneling spectrum (*dI/dV*) of FeSe surface is shown in Fig. 2(b) (blue curve). It displays a full superconducting gap with two pairs of coherence peaks (referred as $\Delta_1$ and $\Delta_2$) and a flat bottom, as observed in bulk crystals [34,35]. Since our band calculations and ARPES measurements revealed surface states at $\bar{\Gamma}$, it is possible that the double gaps actually originate from the bulk and the surface band, as discussed below. For the $(Li_{1-x}Fe_x)OH$ surface, the tunneling spectrum typically shows metallic behavior with a weak dip *off* the Fermi level (green curve in Fig. 2(b)).

To determine the gap of the FeSe surface more precisely, we fit the tunneling spectrum with a Dynes function (see section **S4** of Supplementary Materials for details). The observed gaps are broader than an isotropic gap with a similar size (grey curve in Fig. 2(c)). However, the line shape of the smaller gap ($\Delta_1$) can be well fitted by an empirical anisotropic gap function: $\Delta_1(k)=\Delta_1^{min}+(\Delta_1^{max}-\Delta_1^{min})|\cos2\theta|$ (red curve in Fig. 2(c)) after subtracting a sloping background. The fit yields $\Delta_1^{max}$ = 8.2 meV, $\Delta_1^{min}$ = 5.7 meV and a small Dynes term $\Gamma$ of 0.1 meV. Note that such a $\Delta_1^{max}$ value can also be estimated as half distance between the corresponding two coherence peaks, and so does $\Delta_2^{max}$. However, the larger gap $\Delta_2$ has additional broadening which makes the determination of gap anisotropy difficult. The green curve in Fig. 2(c) gives a simulated spectrum with $\Delta_2^{max}$ = 14.5 meV (see Supplementary Materials for details). We note that the size of $\Delta_1$ (even $\Delta_1^{max}$) is obviously smaller than the ARPES measured gap (~10 meV) on the $\bar{M}$ pocket, thus $\Delta_1$ is unlikely from the $\bar{M}$ pocket, and it can only be attributed to the gap of the topological surface state obtained through proximity from the bulk. On the other hand, $\Delta_2^{max}$ is larger than 10 meV and thus $\Delta_2$ should correspond to the bulk gap on the $\bar{M}$ pocket. As will be shown below, the gap size measured by ARPES reflects the average size of $\Delta_2$.

QPI measurements were carried out on FeSe surface. Figure 2(d) presents a typical fast-Fourier-transformed (FFT) QPI image taken at *E* = 5 meV and *T* = 4.2 K – more data are presented in section **S5** of Supplementary Materials. The dominant features are the ring-like patterns centered at (0, 0), (π, π) and (0, 2π), which arise from the scattering between electron pockets at $\bar{M}$ [34,35]. Figure 2(e) summarizes the averaged FFT line cuts through $q = (\pi, \pi)$, which displays an electron-like dispersion. Using $q = 2k$, a parabolic fit yields a band bottom at -57(±7) meV and $k_F$ = 0.21(±0.02) Å$^{-1}$, consistent with the reported band structure at $\bar{M}$ of $(Li_{1-x}Fe_x)OHFeSe$ [28,29]. In relation to the topological surface states at $\bar{\Gamma}$, we note that a spin-helical structure with spin-momentum locking would strongly suppress backscattering, causing them invisible to QPI.

Under a magnetic field of 10 T perpendicular to the surface, a zero-bias-conductance (ZBC) map is taken on a 36×36 nm$^2$ area of FeSe surface (Fig. 2(f)), as shown in Fig. 2(g). Vortex cores are visible as bright circular regions. We found that many of the vortices are pinned by dimer-like defects, as indicated by the corresponding arrows in Figs. 2(f) and 2(g). The pinned vortices all have a "dark spot"-like feature near their center, which are induced by impurity states. However, we can still find un-pinned or "free" vortices which emerge in defect-free areas, such as the one marked by a dashed circle in Fig. 2(g).

### C. Discrete bound states in the free (un-pinned) vortex cores

The tunneling data of four different free vortices are presented in Fig. 3. Figure 3(a) plots the *dI/dV* line-cut taken across Vortex 1 (inset image, along the arrow); the red curve is measured at the vortex center. It is remarkable that multiple (five) discrete peaks are observed near the core center. There is one peak located exactly at zero bias (a ZBCP), with the other peaks distributed symmetrically around it. The energy spacing between these 5 peaks is close to 1.5 meV. Figure 3(c) presents the spatial evolution of the spectra in a color plot. On leaving the core center, the intensities of these discrete peaks (marked by -2, -1, 0, 1, 2 here, also referred as $E_{-2}$, $E_{-1}$, $E_0$, $E_1$, $E_2$ hereafter) decrease and vanish at ~2nm away. Intriguingly, when the discrete peaks fade out, a pair of much broader peaks (shaded regions in Fig. 3(a)) show up at higher energies. Those peaks keep shifting to higher energy on moving away from the core, giving an "X"-shaped pattern in Fig. 3(c). Figures 3(b) and 3(d) present similar data taken on another free vortex (Vortex 2). Multiple low-energy core states including a ZBCP were also observed, albeit with a slightly smaller energy spacing (~1 meV), and there are also "shifting" high-energy states away from the core center. The behavior of these states, except the one at zero bias, are expected for the CdGM vortex states in quantum limit [13,36].

Such a ZBCP accompanied by discrete CdGM states was repeatedly observed in free vortices. In regions with few defects, such as the one shown in Fig. 3(e), more than one free vortices (Vortex 3 and 4) could be found (Fig. 3(f)). At zero field, a clean superconducting gap is observed over this region, which could vary by about 2 meV along the long cut (Fig. 3(g)). Figures 3(h) and 3(i) show the *dI/dV* line-cuts of Vortex 3 and 4 respectively, focusing on the region close to the vortex center (±1.4 nm) and low energies (±5 meV). There are five discrete peaks with a ZBCP in most spectra, with the energy spacing of 0.8~1.0 meV, well above our energy resolution (0.36 meV).

The positions of $E_{-2}$, $E_{-1}$, $E_1$, and $E_2$ peaks vary in different vortices, and they even vary at different locations in the same vortex. As shown in Figs. 3(h) and 3(i), upon leaving the center, the positions of $E_{-2}$, $E_{-1}$, $E_1$, and $E_2$ peaks shift away from the dashed lines that represent the peak positions at the center. This is because the CdGM states are spaced by $\delta E=\Delta^2/E_F$, and the superconducting gap varies in space (Fig. 3(g)). In contrast, the energy position of the $E_0$ peak is always at zero bias, independent of local gap, which suggests that it is most likely protected by some global properties, such as topology. In addition, since conventional CdGM states are not located at zero energy, but at *half*-integer multiples of $\Delta^2/E_F$, the origin of the $E_0$ peak is highly nontrivial.

The high-level core states are closely packed and form the broad "shifting" peaks. In theory the spacing between high-level states will decrease and make them undistinguishable, while the maximum intensity of core states will shift to high energy on leaving the core center [36-38], resulting in two "splitting" peaks. This behavior has been widely observed, such as in NbSe$_2$ (Ref. [37]); however, individual core states are rarely seen due to the small $\delta E = \Delta^2/E_F$ for most conventional superconductors. Here we clearly observe both discrete low-level core states and the quasi-continuous high-level states, benefitting from the relatively large $\delta E$ of (Li$_{1-x}$Fe$_x$)OHFeSe (as shown below) and sufficiently high resolution at 0.4 K (For comparison, see Supplementary Materials section **S6** for the T=4.2 K data).

The well-separated low-level core states enable quantitative analysis on them. We first applied multiple-Gaussian-peak fitting to the summed low-energy spectra taken near the vortex center (over a ~ ±0.7 nm range to reduce the uncertainty from peak fluctuations). The results are shown in Figs. 4(a)-4(d) for Vortex 1~4, respectively. Here each core state peak corresponds to one Gaussian peak. The fitted peak energies are directly labeled in Figs. 4(a)-4(d), and are also illustrated in Fig. 4(e) with error bars that represent the energy *range* of peak fluctuation. The full widths at half maximum (FWHM) of the Gaussian peaks are illustrated in Fig. 4(e) by colored bars. The detailed values of all the fit parameters are listed in Table SI in Supplementary Materials. From the fitting we can reveal several important facts:

Firstly, the fitted energy of $E_0$ for all the vortices are negligibly small, typically one order of magnitude smaller than the energy resolution (0.36 meV). The fluctuation range of $E_0$ (error bars in Fig. 4(e)) always covers zero. Meanwhile, the FWHM of the $E_0$ peak is in the range of 0.59~0.80 meV, larger than the energy resolution. However, one cannot attribute such an additional broadening (0.23~0.44 meV) to two unresolved conventional CdGM states sitting close to zero bias, assuming they have the energies of $\pm\frac{1}{2}\Delta^2/E_F$, because it would require a superconducting gap with a mean size of 3.6 ~ 5.0 meV (assuming $E_F$ = 57 meV), and the observed local gap size near these vortices are well above that (see Fig. S6 and Table SII, the size of $\overline{\Delta}_1=(\Delta_1^{max}+\Delta_1^{min})/2$ are all around 7 meV). Besides, the FWHM of $E_0$ is obviously narrower than the FWHM of other CdGM states. As discussed below, the gap anisotropy may induce additional broadening but cannot well account for the finite width of $E_0$. Therefore, within our experimental resolution, $E_0$ is a zero-energy state with finite width, whose origin is yet to be explored.

Secondly, the energies of $E_{\pm 2}$ and $E_{\pm 1}$ states are nearly symmetric with respect to the Fermi level (with different values for different vortices). However, the spacing between $E_2$ and $E_1$ is always slightly larger than that between $E_1$ and $E_0$. The ratio of $(E_2-E_{-2})/(E_1-E_{-1})$ falls between 2.1~2.5 for Vortex 1~4 (see Table SII). This is unexpected if we assume all five states are from one single electronic component, since the spacing between neighboring states will decrease on going to higher energy [36,38]. Therefore, these may actually arise from two bands with different gap size and/or band bottom position, as we did observe a double gap $dI/dV$ spectrum and a topological surface state. Meanwhile, the local gap inhomogeneity and anisotropy should also affect the energy of core state for different vortices. Taking into account these effects, we measured the local gap size ($\Delta_1^{max}$, $\Delta_1^{min}$ and $\Delta_2^{max}$) of the region where these vortices emerge (see Fig. S6 and Table SII for details). As $\Delta_1$ is attributed to the topological surface state, and its approximate mean size is $\overline{\Delta}_1=(\Delta_1^{max}+\Delta_1^{min})/2$; considering the Dirac point is at 20 meV below the Fermi level (defined as $E_D$), then it could reasonably account for the $E_0$ and $E_{\pm 2}$ states for most free vortices through the general formula $E=\mu\Delta^2/E_D$ and $\mu = 0, \pm 1$. In Fig. 4(f) we plot $|E_2|=(E_2-E_{-2})/2$ vs. $(\overline{\Delta}_1)^2$ for different vortices (red spots), a linear fit gives $|E_2|=(0.97\overline{\Delta}_1)^2/E_D$. We note there is a more specific calculation on the vortex states of proximity induced superconductivity in topological surface state [39], which gives $E=\pm 0.83\Delta^2/(\Delta^2+E_D^2)^{1/2}$ for the two second lowest states. This formula can estimate $E_{\pm 2}$ states by taken $\Delta \sim 1.06\overline{\Delta}_1$, which is also in between $\Delta_1^{min}$ and $\Delta_1^{max}$ (see Table SII).

For the topologically trivial bulk state around M with $E_F = 57$ meV, its gap is $\Delta_2$. Its lowest-order core states should be at $\pm \frac{1}{2}\Delta_2^2/E_F$. The mean value of $\Delta_2$ is difficult to determine but $\Delta_2^{max}$ can be estimated by the coherence peaks. We found a linear fit of $|E_I|=(E_I-E_{-I})/2$ vs. $(\Delta_2^{max})^2$ yields $|E_I|=\frac{1}{2}(0.72\Delta_2^{max})^2/E_F$ (blue dashed line in Fig. 4(f)). The value of $0.72\Delta_2^{max} \approx 10$ meV is consistent with the ARPES gap size on the band around the M point (Figs. 1(j) and 1(k)), which should be close to $\overline{\Delta}_2$. Although the two data points marked by dashed circles deviate from the fits, they still show a monotonic relation with gap. We speculate this could be due to some local variation of carrier concentration, which is difficult to determine precisely for each vortex. Despite this uncertainty, the most self-consistent understanding is that the $E_0$ and $E_{\pm 2}$ are from the topological surface state while $E_{\pm I}$ are from the trivial bulk band.

Thirdly, the fitted FWHMs of the $E_{\pm 2}$ and $E_{\pm I}$ states vary between 0.8 to 2 meV. We expect that this is at least partially due to the sizable anisotropy of $\Delta_1$ and $\Delta_2$. The broadening of $E_{\pm 2}$ caused by gap anisotropy of $\Delta_1$ (defined by $\alpha=(\Delta_1^{max}-\Delta_1^{min})/(\Delta_1^{max}+\Delta_1^{min})$) can be estimated by $\delta(\Delta_1^2/E_D)=2\alpha(\overline{\Delta}_1)^2/E_D$, yielding an additional broadening in the range of 0.7~1.5 meV for Vortex 1-4. Meanwhile, although the gap anisotropy of $\Delta_2$ is unknown, if we take the same value with $\alpha$ and assume $\overline{\Delta}_2=0.72\Delta_2^{max}$, its contribution to the FWHM of $E_{\pm I}$ is in the range of 0.3~0.6 meV. Then the resulting total broadenings are compatible with the fitted FWHMs. However, the estimated total broadening of $E_{\pm 2}$ (1.1~1.9 meV) is significantly larger than the FWHM of $E_0$, indicative of the special properties of the latter. It is also worth noting that the core states have different but comparable weight, as reflected by the peak area summarized in Table SI. Moreover, the intensities of $E_{\pm 2}$ and $E_{\pm I}$ are asymmetric. This is consistent with the particle-hole asymmetry expected for a superconductor with small $E_F$ [36,40,41].

Figure 4(g) shows the spatial dependence of the intensity of the core states (dI/dV values at the peak energies, extracted from Fig. 3(c) for Vortex 1). The $E_0$ state is more concentrated at the vortex center (especially compared to $E_{\pm 2}$), consistent with the calculated behavior of the vortex state of a $p+ip$ superconductor [42,43]. An exponential fit (red solid curve) gives a decay length of 1.4 nm, which is an estimation of coherence length. Theoretically, the low-level core states should have an additional oscillation with a period of $\lambda_F=2\pi/k_F$ (characterized by Bessel functions [42,43]) besides the exponential decay. Here $k_F$ is 0.21 Å$^{-1}$ for the bulk band and ~0.03 Å$^{-1}$ for the surface band (from QPI and ARPES measurements), which gives $\lambda_F^B = 3.0$ nm and $\lambda_F^S = 21.0$ nm. These are both significantly larger than $\xi$, thus the $\lambda_F$ oscillation cannot be observed here.

### D. The core states of impurity-pinned vortices

A large number of vortices are pinned by dimer-like defects (Figs. 2(f)-2(g)), as the dimer-like defects at the Fe site can locally suppress superconductivity by inducing impurity states. Figures 5(a) and 5(b) show the tunneling spectra measured at $T = 0.4$ K on two typical dimer-like defects under $B = 0$ T and 10 T (or 11 T). Multiple sharp impurity state peaks can be observed at $B = 0$ T (blue curves), and the peak positions can vary for different defects (see Fig. S8 for additional data). In some cases, a zero-bias peak can show up at the defect site in zero field (Fig. 5(b)). At high field, vortices pinned by these two defects were observed in the ZBC map (the insets of Figs. 5(a) and 5(b)). The spectra on the defect sites now display signatures of split impurity states, such as the double peaks marked by arrows in Fig. 5(a) (see also Fig. S8). The zero-bias peak in Fig. 5(b) is also split away from zero bias. This suggests the dimer-like defects are mostly *magnetic* [44]. Figures 5(c) and 5(d) show the tunneling spectra along the arrows in the insets of Figs. 5(a) and 5(b), respectively. There is no zero-bias peak at the vortex center. Upon moving away from the defect site the impurity states decay, while the broad high-energy core state peaks appear (shaded regions in Figs. 5(c)-5(d)). These observations apparently suggest a competition between CdGM and impurity states. We note that the pinned vortices always have a "hole" in their center in the ZBC map, which also indicates that the low-energy core states are strongly suppressed by the magnetic impurity (see also Figs. S5(d-f)). Nevertheless, we note that in some case the zero-bias vortex state recovers a certain distance away from the defect site, as highlighted by the green curve in Fig. 5(d).

### III. DISCUSSION AND CONCLUSION

Our data represent the cleanest and most robust zero-bias mode observed in vortices so far. The robustness is manifested in the following aspects:

(i) It is always present at zero bias in free vortex cores, regardless of variations in the underlying superconducting gap.

(ii) It can survive to high magnetic field due to the short coherence length. The $\xi$ of $(Li_{1-x}Fe_x)OHFeSe$ is significantly smaller than that of $FeTe_xSe_{1-x}$ (~3 nm) [21] and $Bi_2Te_3/NbSe_2$ (~30 nm) [16], which also means a higher probability to observe multiple free vortices and a lower probability of pinning by impurities during vortex motion. These are critical for braiding and fabricating qubits.

(iii) The large superconducting gap and high $T_c$ makes the system robust against temperature fluctuations and relatively easy to study in experiments.

The cleanness is manifested in three aspects:

(i) The ZBCP is unambiguously separated from other low-lying CdGM states and free from impurity effects in free vortices.

(ii) The FeSe layer is intrinsically stoichiometric, so that improved sample quality will allow more free vortices in larger defect-free regions, enabling further manipulation of the MZMs.

(iii) The width of the ZBCP is the narrowest of all the core states.

The proximity effect of bulk superconductivity induces topological superconductivity in chiral topological surface states [1-3,15], which naturally generates a MZM in the vortex core and explains our observation. Nevertheless, in section **S10** of the Supplementary Materials, we use a two-band model to simulate the vortex states with a variety of possible relevant pairing functions (fully gapped to be consistent with our experiment) of the bulk superconducting state, such as pure *s*-wave, *d+id'*, and nodeless *d*-wave, in both the presence and absence of SOC. None of these scenario would lead to a robust ZBCP. Other origins may lead to a ZBCP in tunneling spectrum such as Kondo effect, SIS tunneling can also be excluded, since there is no impurity in free vortex and the tip used here is non-superconducting. Therefore, so far MZM is the most likely origin of the observed clean and robust ZBCP.

To summarize, our experimental and theoretical findings compellingly demonstrate that $(Li_{0.84}Fe_{0.16})OHFeSe$, a heavily electron doped FeSe-based superconductor, is topologically nontrivial. We present the cleanest and most robust zero-energy modes in vortices so far, which enable us to obtain the important properties of MZMs with unprecedented accuracy and reliability, including their spatial distribution, scattering rate, response to magnetic impurities, and tunneling quantum efficiency as compared to conventional CdGM states, which would put strong constraints on theory. Our work thus presents an ideal and practical platform to further study the properties of MZMs, explore their manipulation such as braiding, and construct MZM-based qubits, which opens a new, clear route to rapid progress in both the fundamental understanding and potential applications.

## Acknowledgements

We thank professors Fuchun Zhang, Jiangping Hu, Jing Wang, Dunghai Lee and Darren Peets for helpful discussions. This work is supported by the National Natural Science Foundation of China, National Key R&D Program of the MOST of China (Grant No. 2016YFA0300200, 2017YFA0303004, 2017YFA0303104, 2016YFA0302300 and 2017YFA0303003), National Basic Research Program of China (973 Program) under grant No. 2015CB921700, Science Challenge Project (grant No. TZ2016004), the Strategic Priority Research Program and Key Research Program of Frontier Sciences of the CAS (grant No. QYZDY-SSW-SLH001), and the Fundamental Research Funds for the Central Universities (Grant No. 310421113). The calculations used high performance computing clusters at Beijing Normal University in Zhuhai and the National Supercomputer Center in Guangzhou.

## APPENDIX: EXPERIMENTS AND METHODS

The high-quality single-crystalline superconducting films of $(Li_{0.84}Fe_{0.16})OHFeSe$ were grown on a $LaAlO_3$ substrate by a matrix-assisted hydrothermal epitaxial method, as described in Refs. [30,31]. The full-widths at half-maximum (FWHM) of their X-ray rocking curves are 0.1~0.12 degrees, indicative of their high quality. The thicknesses of different films vary from 100 to 400 nm.

STM measurements were conducted in a UNISOKU cryogenic STM at $T$ = 0.4 K or 4.2 K. The sample was cleaved at 78 K in ultrahigh vacuum with a base pressure of $5 \times 10^{-11}$ Torr and immediately transferred into the STM module. Pt-Ir tips were used after being treated on a clean Au (111) substrate. dI/dV spectra were collected by a standard lock-in technique with a modulation frequency of 741 Hz and a typical modulation amplitude $\Delta V$ of 0.1 mV at 0.4 K and 1.0 mV at 4.2 K.

ARPES data were taken under an ultra-high vacuum of $5\times10^{-11}$ mbar, with a Fermi Instruments discharge lamp (21.2 eV He-Iα light) and a Scienta DA30 electron analyzer. The energy resolution is 6 meV and the angular resolution is 0.3°.

We used fully charge self-consistent density functional theory combined with dynamical mean-field theory (DFT+DMFT) [45] to calculate the electronic structure of (Li, Fe)OHFeSe in the paramagnetic state. The DFT part is based on the linearized augmented plane wave method as implemented in WIEN2K [46]. A Hubbard $U$ of 5.0 eV and Hund's coupling $J$ = 0.8 eV were used in the calculations, consistent with previous calculations [47]. The DMFT quantum impurity problem was solved using the continuous time quantum Monte Carlo method [48] at temperature $T$ = 116K. We used the experimentally determined crystal structure including the internal atomic positions in our calculations [49]. The surface states were calculated through the iterative Green's function method [50], taking into account the renormalization and shifting of the DFT bands due to strong electronic correlation effects. More specific details of the calculation are described in section **S1** of supplementary materials.

# References

[1] X. L. Qi and S. C. Zhang, *Topological Insulators and Superconductors.* Rev. Mod. Phys. **83**, 1057 (2011).

[2] J. Alicea, *New directions in the pursuit of Majorana fermions in solid state systems.* Rep. Prog. Phys. **75**, 076501 (2012).

[3] C. W. J. Beenakker, *Search for Majorana fermions in superconductors.* Annu. Rev. Condens. Matter Phys. **4**, 113 (2013).

[4] S. D. Sarma, M. Freedman, and C. Nayak, *Topologically protected qubits from a possible non-Abelian fractional quantum Hall State.* Phys. Rev. Lett. **94**, 166802 (2005).

[5] C. Nayak, *et al. Non-Abelian anyons and topological quantum computation.* Rev. Mod. Phys. **80**, 1083 (2008).

[6] V. Mourik, *et al. Signatures of Majorana fermions in hybrid superconductor-semiconductor nanowire devices.* Science **336**, 1003 (2012).

[7] A. Das, *et al. Zero-bias peaks and splitting in an Al-InAs nanowire topological superconductor as a signature of Majorana fermions.* Nat. Phys. **8**, 887 (2012).

[8] S. Nadj-Perge, *et al. Observation of Majorana fermions in ferromagnetic atomic chains on a superconductor.* Science **346**, 602 (2014).

[9] H. Zhang, *et al. Quantized Majorana conductance.* Nature **556**, 74 (2018).

[10] J. Alicea, *et al. Non-Abelian statistics and topological quantum information processing in 1D wire networks.* Nat. Phys. **7**, 412 (2011).

[11] S. Vaitiekėnas, *et al. Selective Area Grown Semiconductor-Superconductor Hybrids: A Basis for Topological Networks.* arXiv: 1802.04210 (2018).

[12] F. Krizek, *et al. Field effect enhancement in buffered quantum nanowire networks.* arXiv: 1802.07808 (2018).

[13] C. Caroli, P. G. de Gennes, and J. Matricon, *Bound Fermion states on a vortex line in a type II superconductor.* J. Phys. Lett. **9**, 307 (1964).

[14] G. E. Volovik, *Fermions on quantized vortices in superfluids and superconductors.* Turk. J. Phys. **20**, 693 (1996).

[15] L. Fu, and C. L. Kane, *Superconducting proximity effect and Majorana Fermions at the surface of a topological insulator.* Phys. Rev. Lett. **100**, 096407 (2008).

[16] J. P. Xu, *et al. Experimental detection of a Majorana mode in the core of a magnetic vortex inside a topological insulator-superconductor $Bi_2Te_3/NbSe_2$ heterostructure.* Phys. Rev. Lett. **114**, 017001 (2015).

[17] H. H. Sun, *et al. Majorana zero mode detected with spin selective Andreev reflection in the vortex of a topological superconductor.* Phys. Rev. Lett. **116**, 257003 (2016).

[18] M. Y. Chen, *et al. Superconductivity with twofold symmetry in $Bi_2Te_3/FeTe_{0.55}Se_{0.45}$ heterostructures.* Science Advances **4**, eaat1084 (2018).

[19] S. Sasaki, *et al. Topological superconductivity in $Cu_xBi_2Se_3$.* Phys. Rev. Lett. **107**, 217001 (2011).

[20] R. Tao, *et al. Direct visualization of the nematic superconductivity in $Cu_xBi_2Se_3$.* arXiv:1804.09122 (2018).

[21] D. F. Wang, *et al. Evidence for Majorana bound state in an iron-based superconductor.* arXiv:1706.06074v3 (2018).

[22] P. Zhang, *et al. Observation of topological superconductivity on the surface of an iron-based superconductor.* Science **360**, 182 (2018).

[23] J. X. Yin, *et al. Observation of a robust zero-energy bound state in iron-based superconductor Fe(Te,Se).* Nat. Phys. **11**, 543 (2015).

[24] M. Y. Chen, *et al. Discrete energy levels of Caroli-de Gennes-Matricon states in quantum limit in $FeTe_{0.55}Se_{0.45}$.* Nat. Commun. **9**, 970 (2018).

[25] N. Hao and J. Hu, *Topological phase in the single-layer FeSe*, Phys. Rev. X, **4**, 031053 (2014).

[26] Z. Wang, *et al. Topological nature of the $FeSe_{0.5}Te_{0.5}$ superconductor*, Phys. Rev. B. **92**, 115119 (2015).

[27] G. Xu, *et al. Topological superconductivity on the surface of Fe-based superconductors.* Phys. Rev. Lett. **117**, 047001 (2016).

[28] X. H. Niu, *et al. Surface electronic structure and isotropic superconducting gap in $(Li_{0.8}Fe_{0.2})OHFeSe$.* Phys. Rev. B **92**, 060504(R) (2015).

[29] L. Zhao, *et al. Common electronic origin of superconductivity in (Li,Fe)OHFeSe bulk superconductor and single-layer $FeSe/SrTiO_3$ films.* Nat. Commun. **7**, 10608 (2016).

[30] Y. L. Huang, *et al. Superconducting (Li,Fe)OHFeSe film of high quality and high critical parameters.* Chin. Phys. Lett. **34**, 077404 (2017).

[31] Y. L. Huang, *et al. Matrix-assisted fabrication and exotic charge mobility of (Li,Fe)OHFeSe superconductor films.* arXiv:1711.02920 (2017).

[32] M. Yi, *et al. Observation of universal strong orbital-dependent correlation effects in iron chalcogenides.* Nat. Commun. **6**, 7777 (2015).
[33] M. Q. Ren, *et al. Superconductivity across Lifshitz transition and anomalous insulating state in surface K-dosed (Li$_{0.8}$Fe$_{0.2}$OH)FeSe.* Science Advances **3**, e1603238 (2017).
[34] Y. J. Yan, *et al. Surface electronic structure and evidence of plain s-wave superconductivity in (Li$_{0.8}$Fe$_{0.2}$)OHFeSe.* Phys. Rev. B **94**, 134502 (2016).
[35] Z. Y. Du, *et al. Scrutinizing the double superconducting gaps and strong coupling pairing in (Li$_{1-x}$Fe$_x$)OHFeSe.* Nat. Commun. **7**, 10565 (2016).
[36] N. Hayashi, T. Isoshima, M. Ichioka, and K. Machida, *Low-lying quasiparticle excitations around a vortex core in quantum limit.* Phys. Rev. Lett. **80**, 2921 (1998).
[37] H. F. Hess, *et al. Scanning-tunneling-microscope observation of the Abrikosov flux lattice and the density of states near and inside a fluxoid.* Phys. Rev. Lett. **62**, 214 (1989).
[38] F. Gygi and M. Schlüter, *Self-consistent electronic structure of a vortex line in a type-II superconductor.* Phys. Rev. B **43**, 7609 (1991).
[39] J. Sau, *et al. Robustness of Majorana fermions in proximity-induced superconductors.* Phys. Rev. B **82**, 094522 (2010).
[40] L. Shan, *et al. Observation of ordered vortices with Andreev bound states in Ba$_{0.6}$K$_{0.4}$Fe$_2$As$_2$.* Nat. Phys. **7**, 325 (2011).
[41] T. Hanaguri, *et al. Scanning tunneling microscopy/spectroscopy of vortices in LiFeAs.* Phys. Rev. B **85**, 214505 (2012).
[42] Y. E. Kraus, *et al. Testing for Majorana zero modes in a $p_x+ip_y$ superconductor at high temperature by tunneling spectroscopy.* Phys. Rev. Lett. **101**, 267002 (2008).
[43] L. H. Hu, *et al. Theory of spin-selective Andreev reflection in the vortex core of a topological superconductor.* Phys. Rev. B **94**, 224501 (2016).
[44] W. Li, *et al. Phase separation and magnetic order in K-doped iron selenide superconductor.* Nat. Phys. **8**, 126 (2012).
[45] G. Kotliar, *et al. Electronic structure calculations with dynamical mean-field theory.* Rev. Mod. Phys. **78**, 865 (2006).
[46] P. Blaha, K. Schwarz, G. Madsen, D. Kvasnicka, and J. Luitz, *WIEN2K, An Augmented Plane Wave+Local Orbitals Program for Calculating Crystal Properties.* (Karlheinz Schwarz, Techn. Universität Wien, Austria, 2001).
[47] Z. P. Yin, K. Haule, and G. Kotliar, *Kinetic frustration and the nature of the magnetic and paramagnetic states in iron pnictides and iron chalcogenides.* Nat. Mater. **10**, 932 (2011).
[48] K. Haule, *Quantum Monte Carlo Impurity Solver for Cluster DMFT and Electronic Structure Calculations in Adjustable Base.* Phys. Rev. B **75**, 155113 (2007).
[49] X. F. Lu, *et al. Coexistence of superconductivity and antiferromagnetism in (Li$_{0.8}$Fe$_{0.2}$)OHFeSe.* Nat. Mater. **14**, 325 (2015).
[50] Q. S. Wu, *et al. Wannier Tools: An open-source software package for novel topological materials.* Comput. Phys. Commun. **224**, 405 (2018).

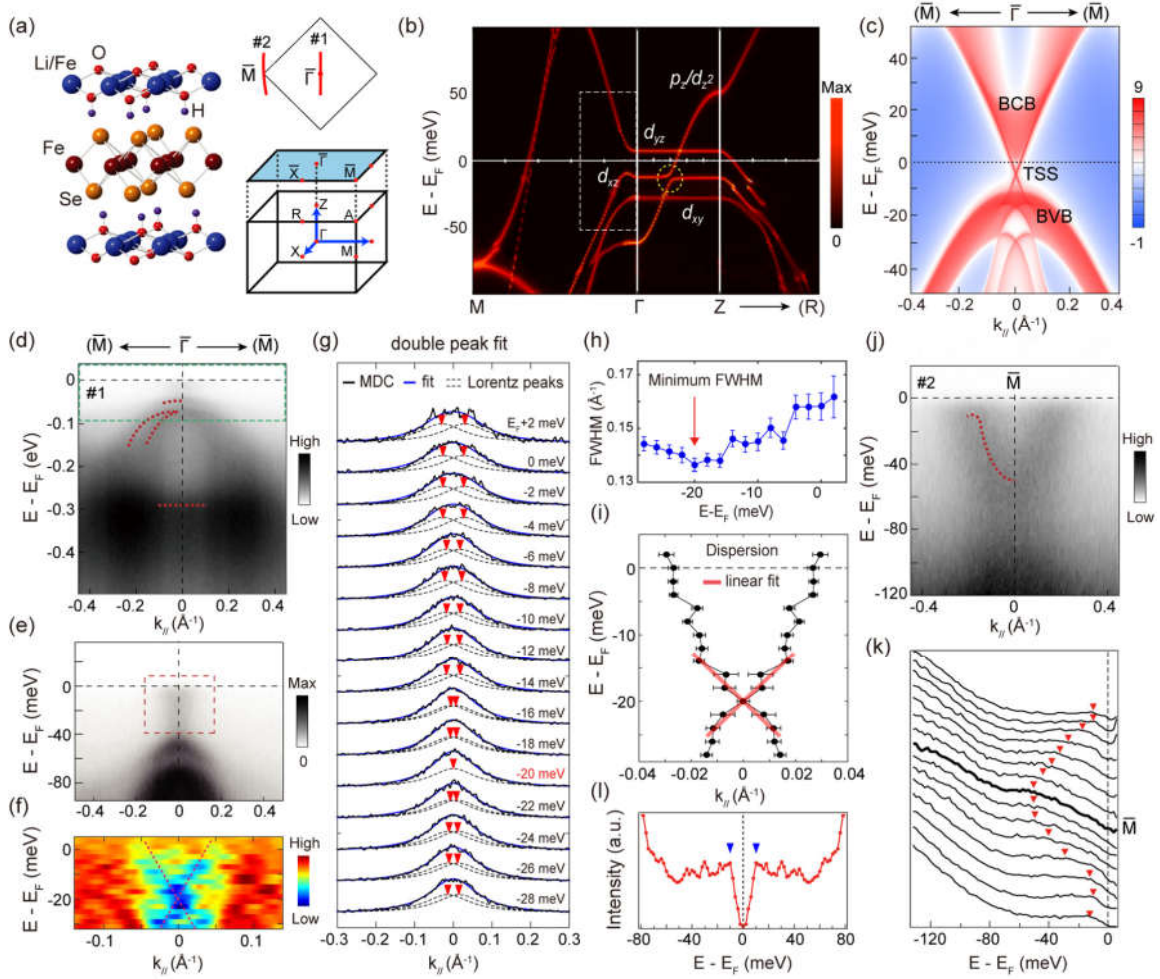

FIG. 1. Band structure and topological surface states of $(Li_{1-x}Fe_x)OHFeSe$. (a) Crystal structure (left), surface Brillouin zone (upper right) and schematic of the projection from 3D to 2D Brillouin zone on (001) surface (lower right) of $(Li_{1-x}Fe_x)OHFeSe$. (b) Band structure of $(Li_{0.75}Fe_{0.25})OHFeSe$ along $M$-$\Gamma$-$Z$-$(R)$ direction, represented by spectral functions calculated by DFT combined with DMFT methods (more extended band structure is shown in Fig. S1(a)). The dashed circle indicates the small SOC gap at the crossing point of Se $4p_z$ and Fe $3d_{xz}$ bands. Dashed rectangle corresponds to left half of the energy/momentum range of panel (c). (c) Calculated bulk and surface states on the (001) surface along the $\bar{\Gamma} - \bar{M}$ direction, illustrated by spectral function. Dirac-cone-like topological surface states (TSS) centered at $\bar{\Gamma}$ connect the bulk valence band (BVB) and conduction band (BCB). The BVB and BCB are mainly composed of $3d_{xz}$ and $3d_{yz}$ orbitals along this direction, respectively. (d) Photoemission intensity of $(Li_{0.84}Fe_{0.16})OHFeSe$ measured across $\bar{\Gamma}$ along cut #1 in panel (a). The dashed curves show the bulk band structure determined by second derivatives. The green dashed rectangle corresponds to the energy/momentum range of panel (e). (e) An enlargement of data in panel (d) near $E_F$ at $\bar{\Gamma}$ point. Finite spectral weight can be clearly observed within the bulk band gap. (f) Second derivative of the photoemission intensity in the marked region of panel (e), a Dirac-cone like dispersion can be seen. (g) Momentum distribution curves (MDCs) of the data in panel (e) in the energy range of -28meV ~ 2meV (normalized by the intensity near $k_{//}$=0). Red markers indicate the Dirac-cone like dispersion extracted from the two-Lorentzian-peak fitting (blue and dashed curves). (h) The FWHM obtained from the one-peak fit (shown in Fig. S2(a)), as a function of energy. Error bars represent 95% confidence bounds. (i) E-k dispersion extracted from the two-peak fit in panel (g), and the linear fit around the crossing point. Error bars represent 95% confidence bounds of the fitted peak position. (j) Photoemission intensity taken along cut #2 across $\bar{M}$ in panel (a). Dashed curve shows the dispersion of the bulk electron band. (k) The energy distribution curves (EDCs) of the data in panel (j) after dividing by Fermi-Dirac distribution. (l) Symmetrized EDC near the Fermi crossing of the $M$ pocket, where a superconducting gap of ~10 meV is observed (indicated by red arrows). All the data were measured at 5.6 K using 21.2 eV photons.

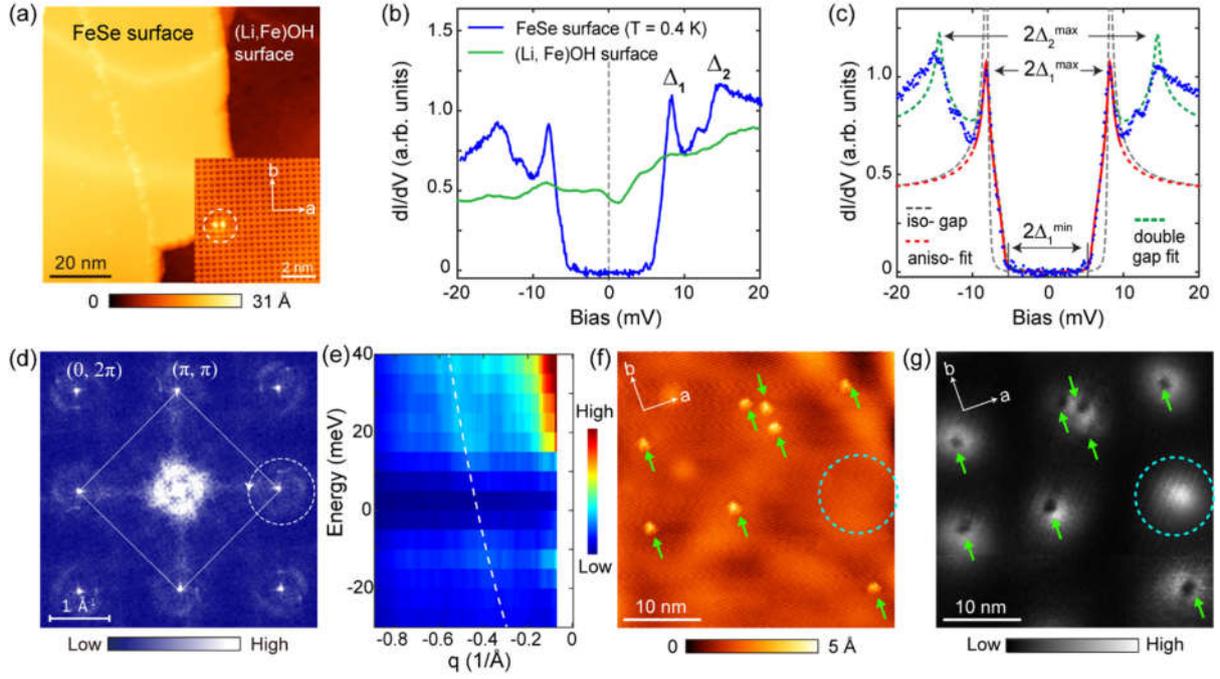

FIG. 2. Topography, superconducting gap, QPI and vortex mapping of $(Li_{0.84}Fe_{0.16})OHFeSe$ film. (a) Topographic image of a cleaved film ($V_b$ = 1 V, $I$ = 5 pA). Inset: the lattice of FeSe surface ($V_b$ = 50 mV, $I$ = 20 pA). A dimer-like defect is marked by the circle. (b) Typical $dI/dV$ spectra taken on FeSe surface at 0.4 K ($V_b$ = -20 mV, $I$ = 60 pA) and on $(Li_{0.84}Fe_{0.16})OH$ surface at 4.2 K ($V_b$ = -20 mV, $I$ = 60 pA). $\Delta_1$ and $\Delta_2$ refer to the smaller and larger gap, respectively. (c) Fits of the superconducting gap (blue dots) to a single isotropic gap function (gray dashed curve), a single anisotropic gap function (red curve), and a double anisotropic gap function (green dashed curve). Details are described in Supplementary Materials. (d) Symmetrized FFT-QPI pattern of an FeSe surface taken at $V_b$ = 5 mV (mapping size: 36 × 36 nm$^2$). (e) Color plot of FFT line cuts through $q = (\pi, \pi)$ (azimuth averaged along the dashed circle in panel (d)). Dashed curve is a parabolic fit to the electron-like dispersion. (f) Topographic image of an FeSe surface ($V_b$ = -30 mV, $I$ = 40 pA, Size: 36 × 36 nm$^2$); green arrows highlight dimer-like defects. (g) ZBC map of the area in panel (f), under $B$ = 10 T. Pinned-vortices are indicated by arrows. The dashed circle encloses a free vortex. If assuming each vortex carries a flux quantum $\Phi_0 = h/2e$, this area should contain $(B \times S)/\Phi_0$ = 6.3 vortices, consistent with the observed number of 7.

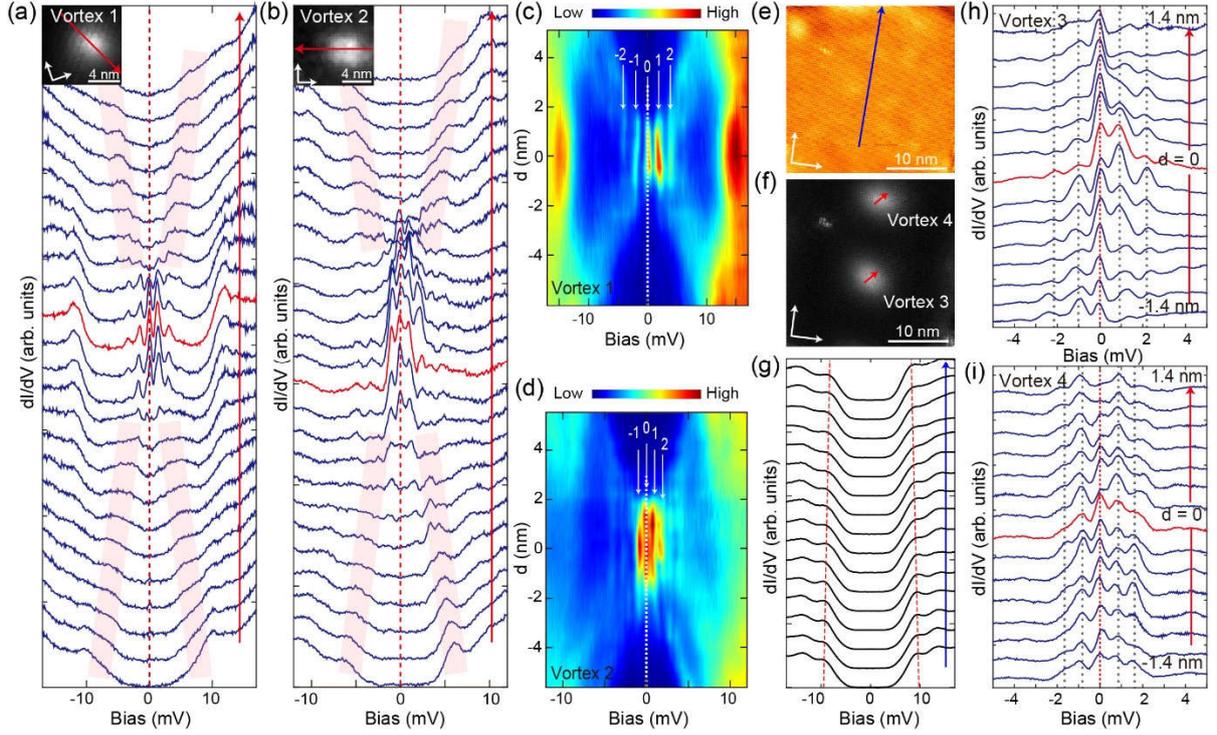

FIG. 3. Bound states in the free vortex cores. (a) and (b) *dI/dV* spectra taken across the free Vortex 1 and 2, shown in the inset images, respectively (setpoint: $V_b$ = -17 mV, $I$ = 60 pA). (c) and (d) Color plots of the spatial dependence of the *dI/dV* spectra shown in panels (a) and (b) respectively. Dashed lines indicate the zero bias, and arrows indicate the positions of discrete low-energy states. (e) and (f) Topographic image and vortex mapping (B= 10 T) of an FeSe surface, respectively. Two free vortices (3 and 4) are observed in this region. (g) *dI/dV* spectra taken along the blue arrow in panel (e), at *T* = 4.2 K and *B* = 0 T. Red dashed lines trace the position of inner coherence peaks. (h) and (i) dI/dV spectra taken along the red arrows in panel (f) across the center of Vortex 3 and 4, respectively (setpoint: $V_b$ = -5 mV, $I$ = 60 pA). Dashed lines indicate peak positions at the vortex center.

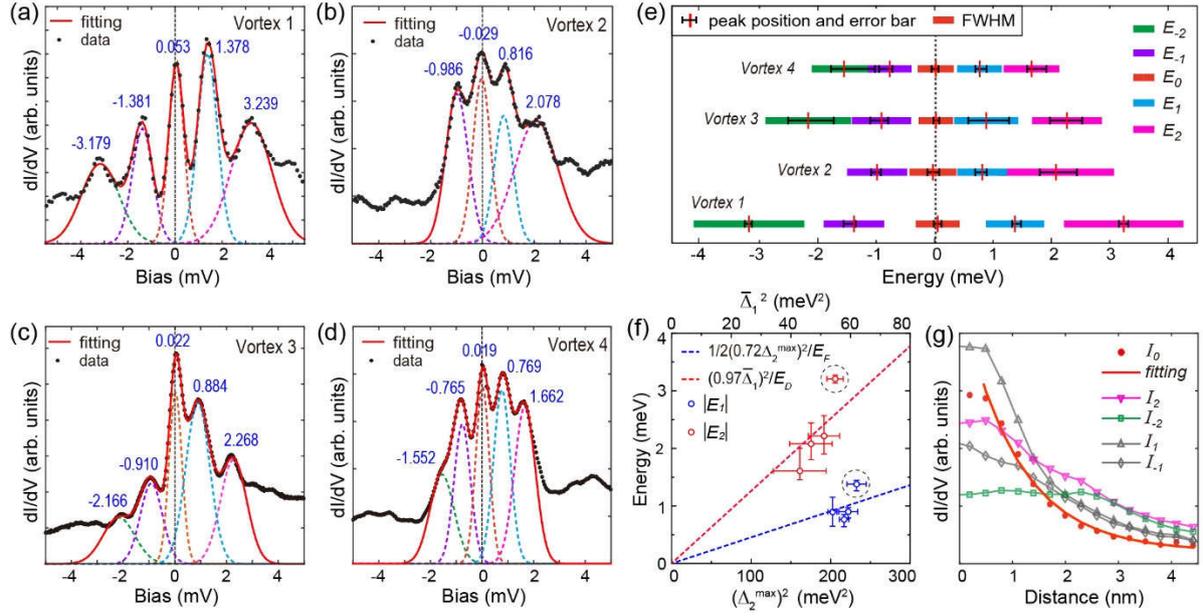

FIG. 4. Quantitative characterization of the vortex core states. (a)-(d) Summed low-energy $dI/dV$ spectra taken near the centers of Vortex 1~4. Black dots are experimental data. Red solid curves are the fits to multiple Gaussian peaks (dashed curves are the individual peaks). (e) Peak energy position (marked by red vertical rods) and FWHM (represented by the length of colored bars) of the core states of Vortex 1~4, obtained from Gaussian peak fitting. The black error bars represent the ranges of peak energy fluctuations (see section **S8** of Supplementary Materials for more details). (f) Plots of $|E_2|$ (red circles) as a function of $(\bar{\Delta}_1)^2$, and $|E_1|$ (blue circles) as a function of $(\Delta_2^{max})^2$. Dashed lines are the linear fitting (see legend). (g) Spatial dependence of the intensity of the core states. $I_2$, $I_1$, $I_0$, $I_{-1}$ and $I_{-2}$ represent the intensity ($dI/dV$ values at the peak energy) of the five core states $E_2$, $E_1$, $E_0$, $E_{-1}$ and $E_{-2}$ of Vortex 1. Red solid curve is the exponential fit to the decay of $E_0$.

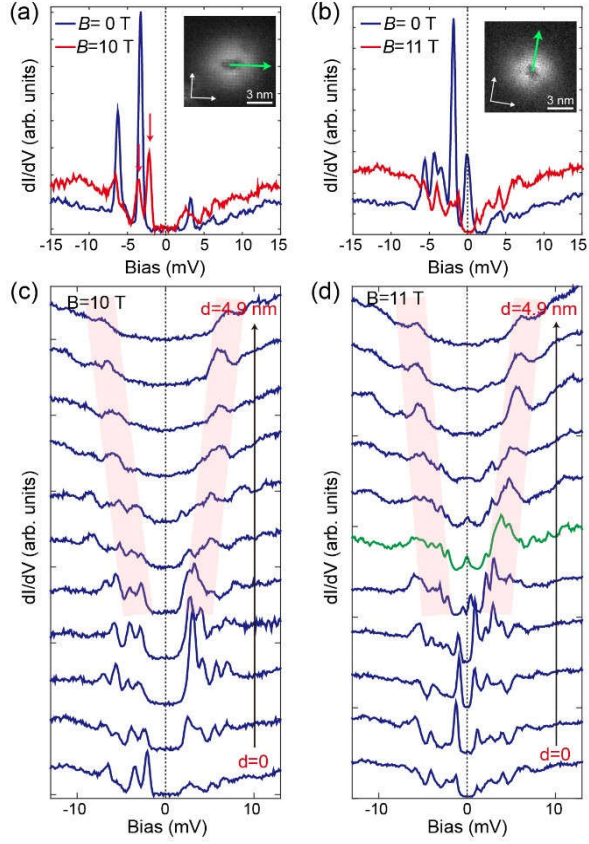

FIG. 5. Bound states in the impurity pinned vortex cores. (a) and (b) *dI/dV* spectra measured on two dimer-like defects at $B = 0$ T and $B = 10$ (11) T ($V_b$ = -17 mV, $I$ = 60 pA). Insets in panels (a) and (b): ZBC maps of the pinned vortex. (c) and (d) *dI/dV* spectra measured along the arrows in the inset image, respectively.

# Supplementary Materials for

# "Robust and clean Majorana zero mode in the vortex core of high-temperature superconductor (Li$_{0.84}$Fe$_{0.16}$)OHFeSe"

Qin Liu[1,2,3+], Chen Chen[1,3+], Tong Zhang[1,3*], Rui Peng[1,3], Ya-Jun Yan[1,3], Chen-Hao-Ping Wen[1,3], Xia Lou[1,3], Yu-Long Huang[4], Jin-Peng Tian[4], Xiao-Li Dong[4], Guang-Wei Wang[5], Wei-Cheng Bao[6,7], Qiang-Hua Wang[3,6], Zhi-Ping Yin[5*], Zhong-Xian Zhao[4], Dong-Lai Feng[1,3*]

[1] State Key Laboratory of Surface Physics, Department of Physics, and Advanced Materials Laboratory, Fudan University, Shanghai 200438, China
[2] Science and Technology on Surface Physics and Chemistry Laboratory, Mianyang, Sichuan 621908, China
[3] Collaborative Innovation Center of Advanced Microstructures, Nanjing 210093, China
[4] Beijing National Laboratory for Condensed Matter Physics and Institute of Physics, Chinese Academy of Sciences, Beijing 100190, China
[5] Department of Physics and Center for Advanced Quantum Studies, Beijing Normal University, Beijing 100875, China
[6] National Laboratory of Solid State Microstructures & School of Physics, Nanjing University, Nanjing, 210093, China
[7] Zhejiang University of Water Resources and Electric Power, Hangzhou 310018, China

[+] These authors contributed equally.
*Email: tzhang18@fudan.edu.cn, yinzhiping@bnu.edu.cn, dlfeng@fudan.edu.cn

## S1. Details on the DFT/DMFT band calculation

To account for the strong electronic correlation in the iron selenide superconductors, we employed density functional theory combined with dynamical mean field theory (DFT+DMFT) [S1,45] to study the electronic structure and topological properties of (Li$_{0.75}$Fe$_{0.25}$)OHFeSe. The density functional theory (DFT) part is based on the full-potential linear augmented plane wave method implemented in Wien2k [46] in conjunction with Perdew–Burke–Ernzerhof form [S2] of the general gradient approximation to the exchange-correlation functional. DFT+DMFT was implemented on top of Wien2k as detailed in Ref. [S3]. The on-site Coulomb interaction between the Fe 3d electrons was parameterized in the rotationally invariant form by a Hubbard U=5.0 eV and Hund's coupling J=0.8 eV, in consistent with previous calculations [S4,S5,47]. An orbital-dependent exact double-counting scheme was used to account for the strong orbital-selective shifting and renormalization of the DFT band structure in iron selenides [S6]. The electronic charge was fully converged on the whole DFT+DMFT density matrix. The DMFT quantum impurity problem was solved by the continuous time quantum Monte-Carlo (CTQMC) method [48,S7] at a temperature T = 116 K. The experimentally determined crystal structure including the internal atomic positions of all the atoms was used in the calculations [49].

After full charge self-consistency of the DFT+DMFT calculations, we obtained the band structure of $(Li_{0.75}Fe_{0.25})OHFeSe$ (Fig. S1(a)) and parameterized the band structure around the Fermi level in terms of tight-binding hopping parameters of Fe 3d and Se 4p orbitals. DFT tight-binding hopping parameters were used as a starting point, and were obtained using the maximally-localized Wannier functions (MLWF) method [S8] implemented in the WANNIER90 code [S9]. The band structure obtained from the MLWF tight-binding model fits well to the first-principles calculated band structures in the range of -2.5 to 1.8 eV relative to the Fermi energy. The DFT tight-binding hopping parameters were renormalized to reproduce the DFT+DMFT band structure in order to obtain the DFT+DMFT tight-binding Hamiltonian. The surface states were calculated through the iterative Green's function method implemented in the WANNIERTOOLS package [50] using the DFT+DMFT tight-binding Hamiltonian.

We have also carried out DFT+DMFT calculations on LiOHFeSe with the same crystal structure as above but with no Fe atoms in the LiOH layer (See Fig. S1(b)). The Fe 3d orbital dominated bands around the Fermi level are similar to that in $(Li_{0.75}Fe_{0.25})OHFeSe$. However, there is no Se $p_z$ orbital derived band that crosses the Fermi level and the Fe $3d_{xy}$, $3d_{xz}$ and $3d_{yz}$ bands. It indicates that there is no topological band inversion along the $\Gamma$ - Z direction. Therefore, LiOHFeSe has a topologically trivial band structure and cannot host topological surface states on the (001) surface. The Fe atoms in the LiOH layer play an important role in inducing the non-trivial band topology of (Li,Fe)OHFeSe and the topological surface states on its (001) surface.

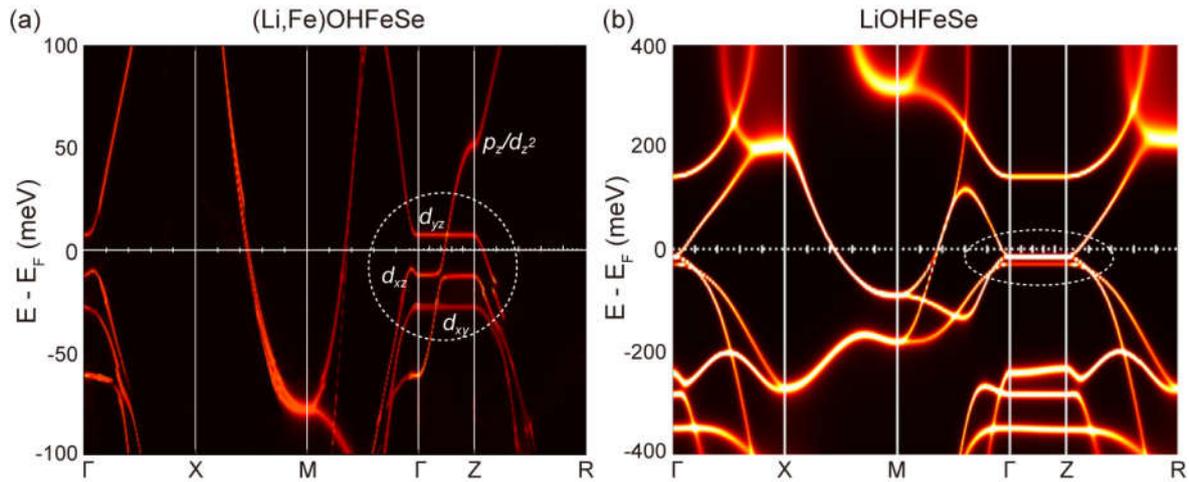

FIG S1. (a) Band structure of $(Li_{0.75}Fe_{0.25})OHFeSe$ along $\Gamma$-X-M-$\Gamma$-Z-R direction. The circle marks the region where the topological band inversion happens. (b) Band structure of LiOHFeSe (i.e., without Fe atoms in the LiOH layer) in the PM state without SOC. The oval marks the same region in panel (a) but there is no Se $4p_z$ band crosses the Fermi level and the Fe 3d bands along the $\Gamma$ - Z direction. No topological band inversion occurs along the $\Gamma$ - Z direction.

## S2. Additional details on the extraction of the surface state dispersion

To explore the band dispersion of surface states, we applied both one and two Lorentzian-peak fittings to the MDCs crossing $\Gamma$, as shown in Figs. S2(a) and (b). The MDCs have been normalized by their intensities near $k_{//}=0$ before the fitting. We found that the peak width (FWHM) in one-peak fitting displays a minimum at $E=-20$ meV, as shown in Fig. 1(h), which would indicate a position of band crossing. In two-peak fitting (Fig. S2(b)), we assume the scattering rate variations are negligible in a small energy window, so that each peak has a constant FWHM as that of the single peak width at $E=-20$ meV. The resulting peak positions then clearly displays a nearly linear, Dirac-cone like dispersion (marked by red arrows). To compare the "goodness" of one- and two-peak fits, in Fig. S2(c) we plot the sum of squares due to error ($\chi^2$) of these fittings. One sees that at the energies away from the crossing point (~ -20 meV), two-peak fit gives lower $\chi^2$ than that of the one-peak fit, which suggests a dispersive band is more appropriate to account for the spectral weight at $\Gamma$.

Then the extracted E-k dispersion with error bars is plotted in Fig. S2(d). At the energies close to crossing point ($\pm5$ meV), we found that a linear fit to dispersion gives lower $\chi^2$ with respect to (two) quadratic fit (as labelled in the figure), suggesting that it is more likely from a Dirac-cone. However, at higher energies the dispersion observably deviates from linear fit and became more quadratic. Since the calculated SOC gap of (Li, Fe)OHFeSe is actually small (~2.5meV), such deviation may be due to the influence of bulk state, despite that in theory the dispersion right at the Dirac point should be always linear. Due to limited energy/momentum resolution here, a more precise determination of the whole surface state dispersion will require further study. Here the linear fit near the crossing point gives $E_D= 20(\pm2)$ meV and $v_F = 5.5(\pm1.5)\times10^4$ m/s. The $k_F$ is determined to be ~0.028 Å$^{-1}$ directly through the MDC near $E_F$.

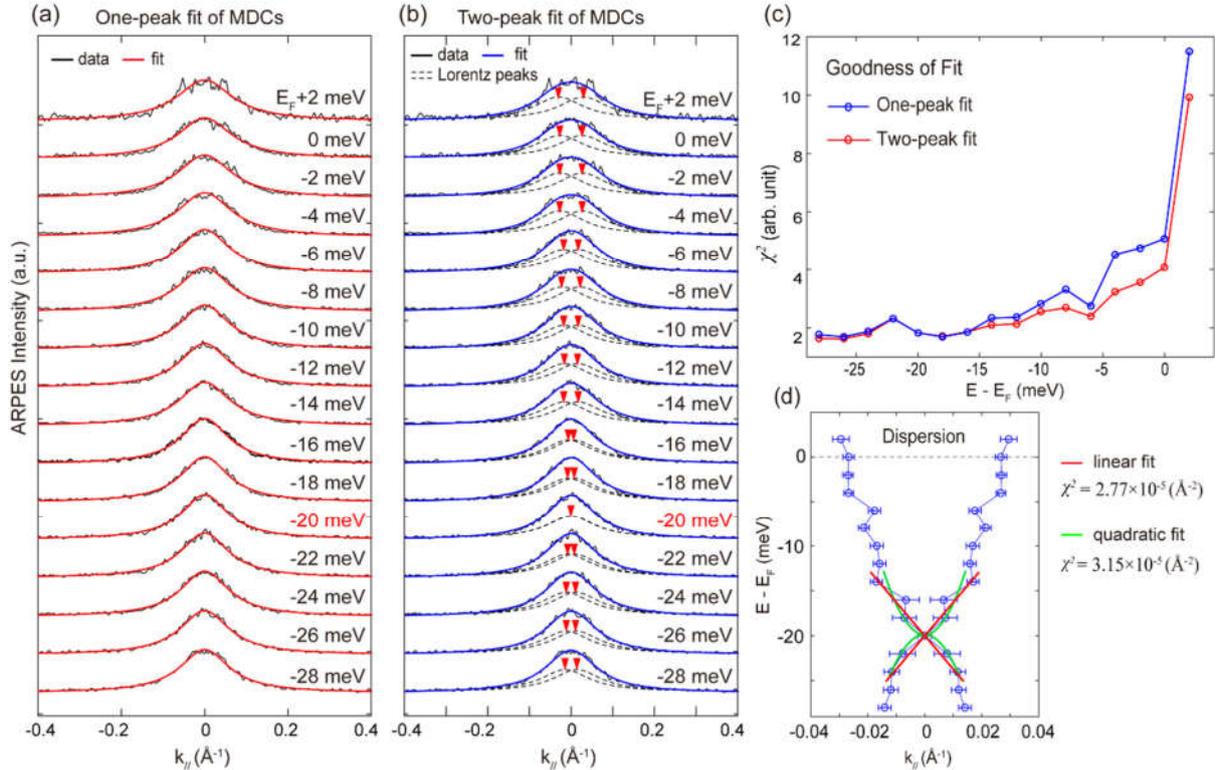

FIG. S2. (a) One-Lorentzian peak fit to the MDC curves near $\Gamma$ (the MDCs have been normalized by their intensities near k=0). (b) Tow-Lorentzian peak fit to the MDCs near $\Gamma$, red markers indicate individual peak positions. (c) The goodness of one and two-peak fits for MDCs at various energy, reflected by the sum of squares due to error ($\chi^2$). At the energies away from the band crossing point (~ -20 meV), the two-peak fit is better than one-peak fit. (d) The E-k dispersion extracted from the two-peak fit (error bars represent the 95% confidence bounds of the fitting parameter). Red and green curves are linear and two quadratic fits to the dispersion near the band crossing point, respectively. Linear fit has a smaller $\chi^2$ than that of the quadratic fit.

## S3. Calibration of STM energy resolution at T = 0.4 K and bias voltage offset

The energy resolution of an STM is limited by thermal and electrical noise broadening. The total broadening can be estimated as $3.5k_BT_{eff}$, where $T_{eff}$ is the effective electron temperature. To check $T_{eff}$, we measured the superconducting gap of a Pb/Si(111) film at $T = 0.4$ K, as shown in Fig. S3(a). A standard BCS fit gives $\Delta = 1.39$ meV, $T_{eff} = 1.18$ K and a small Dynes term $\Gamma = 0.005$ meV which accounts for a finite quasi-particle lifetime. The total broadening (defined as the energy resolution in the text) of the STM is then given by $3.5k_BT_{eff} = 0.36$ meV.

The STM bias voltage applied to the sample usually has a small offset. This offset can be calibrated by measuring I-V curves at different setpoints (tunneling resistance), because all the I-V curves should intersect at a single point where $V = 0$ and $I = 0$. Figure S3(b) shows such a calibration performed on the metallic (Li,Fe)OH surface which has a more linear I-V curve, giving an accurate determination of the bias offset of -0.093(±0.001) meV. All tunneling spectra presented in this paper have been corrected for this bias offset.

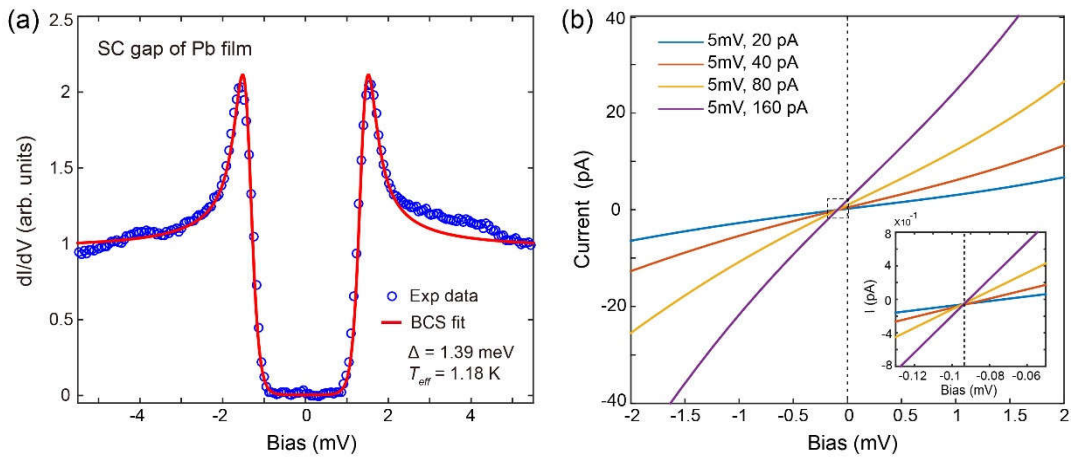

FIG. S3. (a) The superconducting gap of the Pb/Si(111) film (blue dashed circles) measured at $T = 0.4$ K. The red curve is the BCS fit with $\Delta = 1.39$ meV, $T_{eff} = 1.18$ K and $\Gamma = 0.005$ meV. The energy resolution is given by $3.5k_BT_{eff} = 0.36$ meV. (b) A set of I-V spectra taken at different setpoints on the $(Li_{0.84}Fe_{0.16})$OH surface. The intersection point of these I-V curves is the true zero-bias point. Inset: Expanded view of the intersection point.

## S4. Fitting of the superconducting gap of (Li$_{1-x}$Fe$_x$)OHFeSe

The double gapped tunneling spectrum of (Li$_{1-x}$Fe$_x$)OHFeSe cannot be fitted by an isotropic gap (or two), as shown by grey dashed curve in Fig. 2(c). Taking into account the gap anisotropy, the smaller gap $\Delta_1$ can be well fitted by a gap function of: $\Delta_1(k) = \Delta_1^{min} + (\Delta_1^{max} - \Delta_1^{min})|\cos(2\theta_k)|$, where $\Delta_1^{max}$ and $\Delta_1^{min}$ are the maximum and minimum size of $\Delta_1$ in $k$-space. The superconducting DOS is given by the Dynes formula [S10]:

$$N(E)_k = \left|\mathrm{Re}\left(\frac{E-i\Gamma}{\sqrt{(E-i\Gamma)^2 - \Delta_k^2}}\right)\right|$$

The total tunneling conductance is given by:

$$\frac{dI}{dV} \propto \int N(E)_k f'(E+eV)\, dk dE$$

where $f'(E)$ is the derivative of the Fermi-Dirac function at $T_{eff} = 1.18$ K. Before fitting, a linear background is subtracted from the original $dI/dV$ to reduce the line-shape asymmetry (blue dots in Fig. 2(c)). The fitting yields $\Delta_1^{max} = 8.2$ meV, $\Delta_1^{min} = 5.7$ meV, and $\Gamma = 0.1$ meV. The larger gap $\Delta_2$ is much broader than $\Delta_1$ and even an anisotropic gap function cannot give a satisfying fit. The green dashed curve in Fig. 2(c) is a simulated curve using the combined gap function $\Delta(k) = \Delta_1^{min} + (\Delta_1^{max} - \Delta_1^{min})|\cos(2\theta_k)| + A[\Delta_2^{min} + (\Delta_2^{max} - \Delta_2^{min})|\cos(2\theta_k)|]$, where A is the relative weight between the two gaps. The parameters are chosen as $\Delta_1^{max} = 8.2$ meV, $\Delta_1^{min} = 5.7$ meV, $\Delta_2^{max} = 14.5$ meV, $\Delta_2^{min} = 8.5$ meV, $\Gamma = 0.1$ meV and A = 0.8. It is seen that $\Delta_2$ has additional broadening with comparing to simulated curve, which indicates that the anisotropy of $\Delta_2$ cannot be precisely derived from fitting. However $\Delta_2^{max}$ can still be estimated as half the distance between two outer coherence peaks, as the simulated curve always peaks at the maximum value of the gap.

## S5. Additional QPI Results

Additional $dI/dV$ maps and FFT images are shown in Fig. S4.

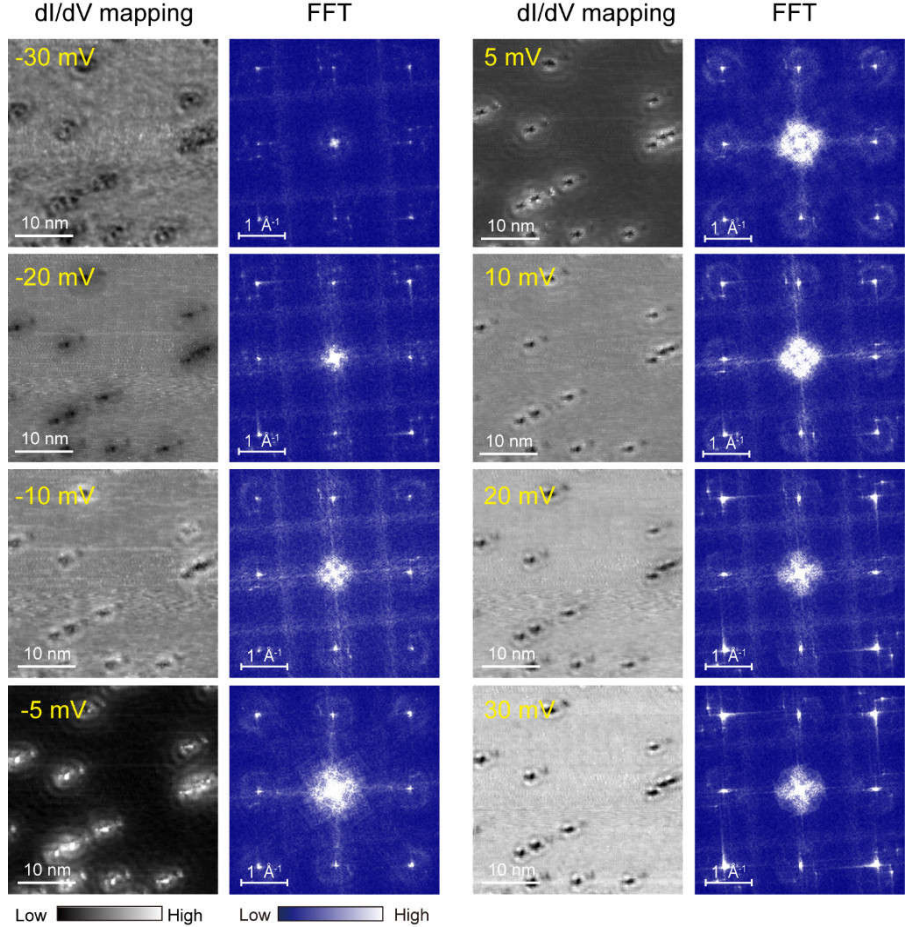

FIG. S4. *dI/dV* maps taken on a 36 × 36 nm$^2$ area of the FeSe surface, and the corresponding FFT images (four-fold symmetrized). The energy for each map is labeled in the image. Each map has 300 × 300 pixels.

## S6. The vortex core states of free and pinned vortices measured at *T* = 4.2 K

Figs. S5(a)-S5(c) show *dI/dV* spectra taken across the same free vortex in the Fig. 3(a) inset, but measured at *T* = 4.2 K. At this elevated temperature, only a single broad peak near zero bias can be observed at the core center, and it "splits" upon moving away from the core center. This is due to enhanced thermal broadening at 4.2 K, which smears out the discrete low-level core states. As the most likely consequence, the zero bias mode was not resolved in our early STS work on Li$_{1-x}$Fe$_x$OHFeSe (Ref. [34]).

Figs. S5(d)-S5(f) show *dI/dV* spectra taken across a pinned vortex, measured at *T* = 4.2 K. One can clearly see from the color plot that the low-energy core states are gapped out by the impurity at the core center.

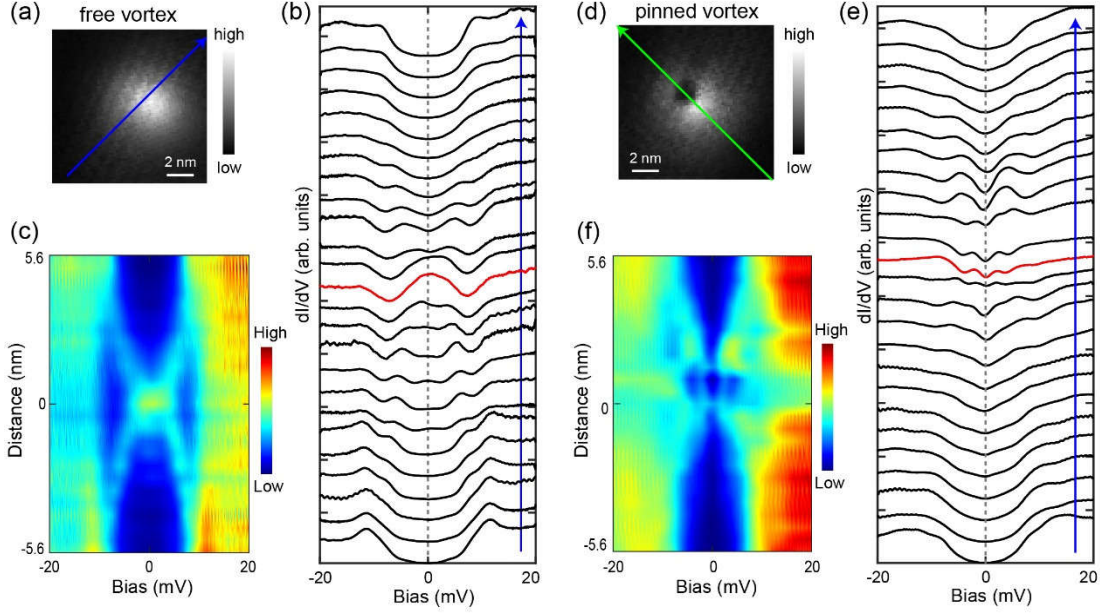

FIG. S5. The vortex core states measured at $T = 4.2$ K. (a), (d) ZBC mapping of an FeSe-terminated surface including a free vortex (panel (a)) and a pinned one (panel (d)) at $B = 10$ T. (b), (e) A series of $dI/dV$ spectra taken across the free and pinned vortex core, respectively, along the arrows in (a) and (d) ($T = 4.2$ K, $V_b = -20$ mV, $I = 60$ pA). (c), (f) Color plots of the $dI/dV$ spectra shown in (b) and (e), respectively.

## S7. Local superconducting gap of the regions where free vortices emerge

We measured the local superconducting gap in the regions near Vortex 1~4, to exclude any impurity state in these regions and obtain the local gap size. Figure S6(a) summarizes the dI/dV spectra measured near the center position of Vortex 1~4 (at $T = 4.2$ K). All the spectra display a clean superconducting gap with double coherence peaks and no in-gap states. To extract the values of $\Delta_1^{max}$ and $\Delta_1^{min}$, we fit the "inner" gap of each spectrum using the same method described in section **S4**, as shown in Figs. S6(b)-S6(e). The values of $\Delta_2^{max}$ are obtained by measuring the half distance between two outer coherence peaks in each spectrum. The value of $\Delta_1^{max}$, $\Delta_1^{min}$ and $\Delta_2^{max}$ for Vortex 1~4 are summarized in Table SII.

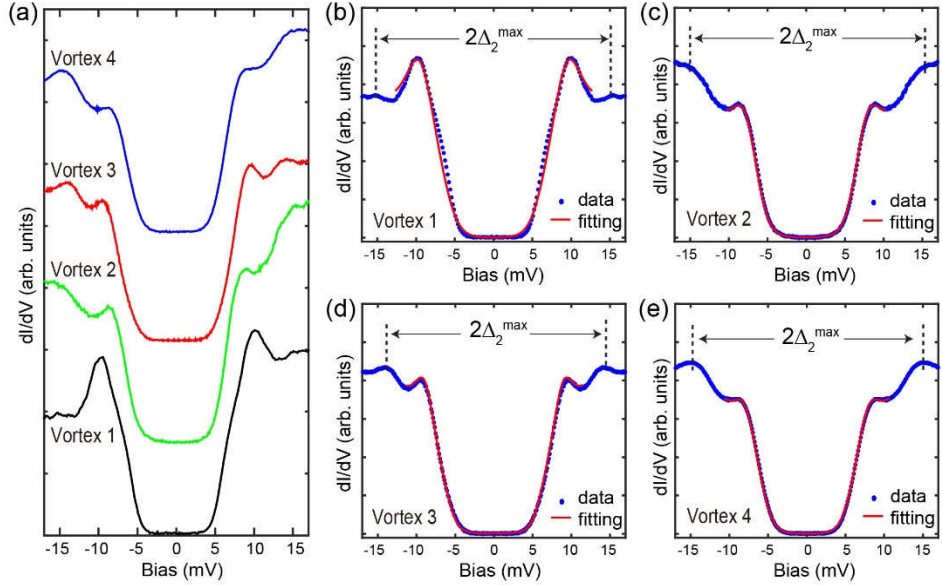

FIG. S6. Local superconducting gaps measured where Vortex 1~4 emerge. (a) Typical dI/dV spectra measured near the centers of Vortex 1~4 at $T$ = 4.2 K and $B$ = 0 T ($V_b$ = -17 mV, $I$ = 60 pA). (b)-(e) Red curves: fits for the inner superconducting gap $\Delta_1$ measured near the center position of Vortex 1~4, using the anisotropic gap function $\Delta_1(k)=\Delta_1^{min}+(\Delta_1^{max}-\Delta_1^{min})|\cos(2\theta_k)|$. Blue dots are symmetrized dI/dV spectra after subtracting a background slope, to reduce the asymmetry of the line-shape of the gap. The energy positions of the outer coherence peaks, which represent $\Delta_2^{max}$, are marked by two dashed vertical lines in each panel. The values of $\Delta_1^{max}$, $\Delta_1^{min}$ and $\Delta_2^{max}$ for Vortex 1~4 are listed in Table SII.

## S8. Quantitative characterization of the vortex core states

As described in the main text, we fit the summed spectra near the centers of Vortex 1~4 with multiple Gaussian peaks. The fitting curves are shown in Figs. 4(a)-4(d) and the fit parameters are summarized Table SI, including the peak position, FWHM and peak area. Uncertainties are listed following the main value. We note the energies of the core states exhibit spatial fluctuations, which is likely due to variation of local gap size (as shown in Figs. 3(h)-3(i)). To characterize this uncertainty, we performed similar Gaussian-peak fitting to *each* individual spectrum taken near the core center of Vortex 1~4 (typically within ±0.7 nm of the core center). The yielded peak positions at different sites reflect their spatial fluctuation. The total range of peak fluctuation are shown in Table SI (in parentheses), and correspond to the error bars in Fig. 4(e) (the uncertainties in the fit peak positions are negligibly small compared to the fluctuation range). In Fig. S7 we plot all the fitted peak positions measured at different sites near the core center.

To clarify the origin of the five core states, we studied the correlation between the core state energies and the local superconducting gap measured near each free vortex. In Table SII, we summarized the related parameters of Vortex 1~4 including: $|E_1| = (E_1 - E_{-1})/2$, $|E_2| = (E_2 - E_{-2})/2$; $|E_2|/|E_1|$, $\Delta_1^{min}$, $\Delta_1^{max}$ and anisotropic ratio α of $\Delta_1$ defined by $\alpha=(\Delta_1^{max}-\Delta_1^{min})/(\Delta_1^{max}+\Delta_1^{min})$ ($\Delta_1^{min}$, $\Delta_1^{max}$ are determined from the fit in Fig. S6), $\Delta_2^{max}$ (determined as half the distance between the two outer coherence peaks); the calculated value of $E_2=(0.97\overline{\Delta}_1)^2/E_D$ or

$E_2=0.83(\Delta_1')^2/\sqrt{(\Delta_1')^2+E_D^2}$ (where $\Delta_1'=1.06\overline{\Delta}_1$), and $E_1=\frac{1}{2}(0.72\Delta_2^{max})^2/E_F$, which basically match the values of $|E_2|$ and $|E_1|$ for most vortices (except Vortex 1).

TABLE SI. Fit parameters of the core state peaks of Vortex 1~4 (Unit: meV).

|  |  | $E_{-2}$ | $E_{-1}$ | $E_0$ | $E_1$ | $E_2$ |
|---|---|---|---|---|---|---|
| Vortex 1 | Peak position[a] | -3.179 ±0.061 (-3.253~-3.130) | -1.381 ±0.031 (-1.561~-1.352) | 0.053 ±0.013 (-0.012~0.118) | 1.378 ±0.019 (1.326~1.483) | 3.239 ±0.051 (3.159~3.318) |
|  | FWHM | 1.897 ±0.262 | 1.036 ±0.080 | 0.753 ±0.036 | 0.999 ±0.052 | 2.045 ±0.253 |
|  | Peak area[b] | 1.118 | 0.882 | 1 | 1.399 | 1.823 |
| Vortex 2 | Peak position |  | -0.986 ±0.039 (-1.088~-0.918) | -0.029±0.027 (-0.121~0.079) | 0.816±0.043 (0.695~0.893) | 2.078±0.067 (1.804~2.439) |
|  | FWHM |  | 1.023±0.082 | 0.803±0.093 | 0.852±0.126 | 1.998±0.405 |
|  | Peak area |  | 1.156 | 1 | 0.822 | 1.846 |
| Vortex 3 | Peak position | -2.166±0.077 (-2.507~-1.729) | -0.910±0.030 (-1.103~-0.795) | 0.022±0.005 (-0.018~0.080) | 0.884±0.010 (0.578~1.281) | 2.268±0.017 (1.976~2.530) |
|  | FWHM | 1.461±0.294 | 1.015±0.083 | 0.586±0.015 | 1.099±0.048 | 1.199±0.065 |
|  | Peak area | 0.672 | 0.813 | 1 | 1.761 | 1.245 |
| Vortex 4 | Peak position | -1.552 ±0.086 (-1.776~-1.018) | -0.765 ±0.023 (-0.955~-0.719) | 0.019 ±0.006 (-0.054~0.067) | 0.769 ±0.010 (0.703~0.888) | 1.662 ±0.016 (1.587~1.919) |
|  | FWHM | 1.105 ±0.244 | 0.750 ±0.056 | 0.619 ±0.021 | 0.766 ±0.039 | 0.951 ±0.056 |
|  | Peak area | 0.900 | 0.954 | 1 | 1.214 | 1.363 |

[a] Numbers in parentheses are the fluctuation range of peak positions.
[b] Peak areas are normalized by the $E_0$ peak of each free vortex.

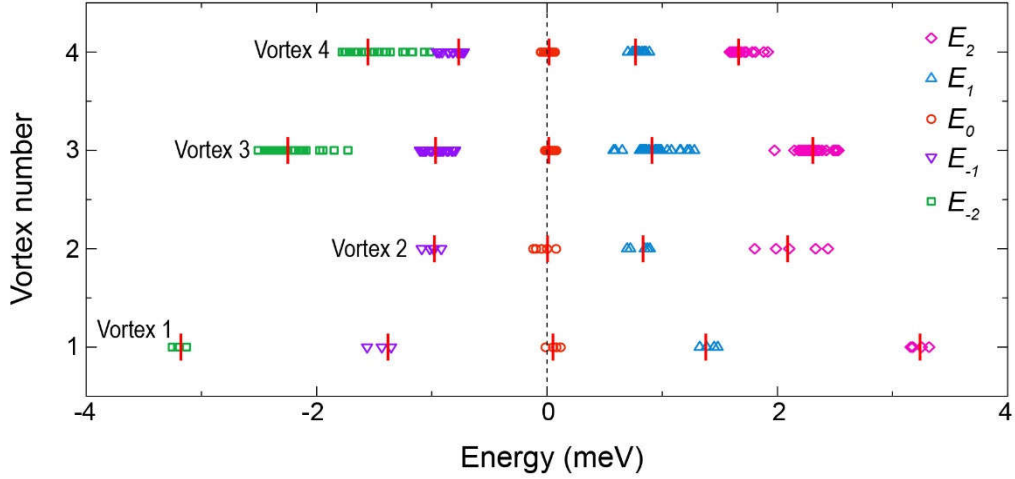

FIG. S7. Peak positions obtained from Gaussian fits to all the measured spectra near the core centers of Vortex 1~4.

TABLE SII. Relation between the core state energy and local SC gap size (Unit: meV).

| Vortex | $|E_2|$ | $|E_1|$ | $\frac{|E_2|}{|E_1|}$ | $\Delta_1^{max}$ | $\Delta_1^{min}$ | $\overline{\Delta}_1$ | $\Delta_1' = 1.06\overline{\Delta}_1$ | $\alpha$ | $\Delta_2^{max}$ | $0.72\Delta_2^{max}$ | $E_2 = \frac{(0.97\overline{\Delta}_1)^2}{E_D}$ | $E_2 = \frac{0.83(\Delta_1')^2}{\sqrt{(\Delta_1')^2 + E_D^2}}$ | $E_1 = \frac{(0.72\Delta_2^{max})^2}{2E_F}$ |
|---|---|---|---|---|---|---|---|---|---|---|---|---|---|
| 1 | 3.21 | 1.38 | 2.33 | 9.40 | 5.41 | 7.40 | 7.84 | 0.27 | 15.25 | 10.97 | 2.58[b] | 2.38[b] | 1.06[b] |
| 2 | 2.08[a] | 0.90 | 2.31 | 7.87 | 5.80 | 6.84 | 7.25 | 0.15 | 14.90 | 10.73 | 2.20 | 2.05 | 1.01 |
| 3 | 2.22 | 0.90 | 2.47 | 8.90 | 5.40 | 7.15 | 7.58 | 0.24 | 14.23 | 10.24 | 2.41 | 2.23 | 0.92 |
| 4 | 1.61 | 0.77 | 2.09 | 7.90 | 5.21 | 6.56 | 6.95 | 0.20 | 14.73 | 10.61 | 2.02 | 1.89 | 0.99 |

[a] Here $|E_2|=E_2$ for Vortex 2.
[b] These values do not well estimate the measured value of $E_2$ and $E_1$ of Vortex 1, which could be due to the local variation of $E_F$, as discussed in the main text. However the measured value of $E_2$ and $E_1$ still show a monotonic relation with gap size.

## S9. Additional data on impurity states induced by dimer-like defects and their response to high magnetic field

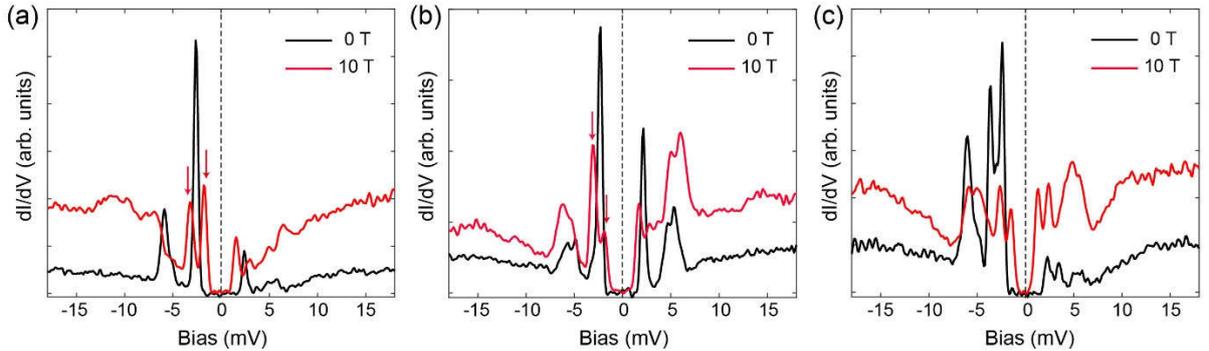

FIG. S8. Additional data on impurity states induced by dimer-like defects, measured at $B$ = 0 T and 10 T. (a)-(c) Comparison of $dI/dV$ spectra measured at $B$ = 0 T and $B$ = 10 T for three native dimer-like defects observed on FeSe-terminated surfaces at $T$ = 0.4 K ($V_b$ = -20 mV, $I$ = 60 pA). Arrows mark the split impurity state peaks at high field.

## S10. Vortex states in a two-band model of (Li$_{1-x}$Fe$_x$)OHFeSe

The behavior of vortex states in a superconductor depends strongly on its pairing symmetry. Considering the vast amount of existing data on heavily electron-doped FeSe-based superconductors (HEFBS), such as K$_x$Fe$_2$Se$_2$, FeSe/STO and (Li$_{1-x}$Fe$_x$)OHFeSe, the Cooper pairs in their bulk states are most likely to be singlet. However, a consensus has not been reached. An early scanning tunneling spectroscopy (STS) study suggested that (Li$_{1-x}$Fe$_x$)OHFeSe may host pure $s$-wave pairing, mainly based on non-magnetic impurities (Zn) not inducing in-gap states (Ref. [34]), whereas a recent phase-sensitive QPI study suggested a sign-reversal pairing within the electronic pockets for (Li$_{1-x}$Fe$_x$)OHFe$_{1-y}$Zn$_y$Se [S11]. An inelastic neutron scattering study suggested a remarkable transition from sign-reversed to sign-preserved pairing symmetry for (Li$_{0.8}$Fe$_{0.2}$)ODFeSe [S12]. Below we simulate the vortex core states of (Li$_{1-x}$Fe$_x$)OHFeSe with a variety of possible (bulk) pairing symmetries, using a two-band model.

The bulk electronic structure of (Li$_{1-x}$Fe$_x$)OHFeSe has two electron pockets at $M$. Before they are folded into the reduced 2Fe/cell Brillouin zone, they can be viewed as a pocket around X and another around Y in the 1Fe/cell Brillouin zone. Upon folding, the two pockets intersect – see Fig. S9(a). Following [S13] we adopt a $\boldsymbol{k\cdot p}$ model and compactify it on a lattice. This is sufficient to describe the low energy quasiparticles. The normal-state single-particle Hamiltonian is, in momentum space,

$$h_{\boldsymbol{k}} = \epsilon_{\boldsymbol{k}} + d_{\boldsymbol{k}}\sigma_3 + \boldsymbol{g}_{\boldsymbol{k}}\cdot\boldsymbol{s}\,\sigma_1$$

Henceforth, $\sigma_{1,2,3}$ are Pauli matrices acting on the two effective orbitals, and $s_{1,2,3}$ are Pauli matrices acting on spins. Agterburg et al proposed, in the continuum limit [S13]:

$$\epsilon_{\boldsymbol{k}} = \frac{k^2}{2m} - E_f, \quad d_{\boldsymbol{k}} = \alpha k_x k_y, \quad \boldsymbol{g}_{\boldsymbol{k}} = \beta(k_y, k_x)$$

where $m$ is the effective mass, and $\alpha$ and $\beta$ are coefficients. The above form of $h_{\boldsymbol{k}}$ takes proper account of the symmetry of the effective orbitals at the $M$ point near the Fermi level. We rotate the coordinate frame (by 45º about $z$) and spin axis independently, so that

$$\epsilon_{\boldsymbol{k}} \to \frac{k^2}{2m} - E_f, \quad d_{\boldsymbol{k}} \to \alpha'(k_x^2 - k_y^2), \quad \boldsymbol{g}_{\boldsymbol{k}} \to \beta'(k_y, k_x)$$

with modified coefficients. Notice that rotation of the spin axis by constant Euler angles does not alter the singlet pairing. The advantage of the rotated $h_{\boldsymbol{k}}$ is an emergent superficial $\boldsymbol{C_{4v}}$ symmetry, with the two orbitals behaving effectively as $xz$ and $yz$. (Notice that the resulting Fermi pockets are elongated along $x$ and $y$ in the new frame.) We then compactify the model on a lattice, with

$$\epsilon_{\boldsymbol{k}} \to -2t(\cos k_x + \cos k_y) - \mu, \quad d_{\boldsymbol{k}} \to 2t'(\cos k_x - \cos k_y), \quad \boldsymbol{g}_{\boldsymbol{k}} \to 2\lambda(\sin k_y, \sin k_x)$$

where $t'$ accounts for the Fermi pocket anisotropy, and $\lambda$ measures the strength of the spin-orbital coupling (SOC). We set $t = 1$, $\mu = -3.5t$ (or $E_f = 0.5t$) to have shallow electron pockets mimicking the experimental situation. The small parameters $t'$ and $\lambda$ will be specified in due course. A further advantage of the rotated $h_k$ arises after the compactification: the symmetric hopping integrals, the $d$-wave like anisotropy and the SOC can all be defined on nearest-neighbor bonds.

In the superconducting state, the vortex bound states would be the usual Caroli-de Gennes-Matricon states for $s$-wave pairing. Here we first consider a less trivial case, so-called nodeless $d$-wave pairing. In such a case, there is a full gap on both pockets, but with a sign change from X to Y (see Fig. S9(a)). One may worry that upon hybridization, the energy gap on the reconstructed bands, illustrated in Fig. S9(b), would have to be nodal. However, this does not have to be so if the hybridization is from SOC, as shown in Ref. [S13]. We now ask how the nodeless $d$-wave would impact on the vortex bound states. We write the pairing part of the Hamiltonian as, in momentum space (for the uniform case),

$$\Delta_k = [\Delta_1 \sigma_3 + \Delta_2 (\cos k_x - \cos k_y)] i s_2$$

where $\Delta_1$ is the onsite part, with opposite phase on the two bands before reconstruction, see Fig. S9(a) for illustration, $\Delta_2$ is the amplitude of $d$-wave pairing on nearest bonds, and $i s_2$ is the spin antisymmetric tensor accounting for singlet pairing (note the definitions of $\Delta_1$ and $\Delta_2$ here are different from those in the main text). Note that $\sigma_3$ transforms as $d$-wave under rotation, hence both components in the gap function behave as $d$-wave. The resulting pairing gap is nodeless if $|\Delta_1| > 4|\Delta_2|$. Since $\Delta_2$ leads to gap variation on the Fermi surface, which is weak in experiments (about 10% from ARPES), we shall ignore $\Delta_2$ for the moment (and we verified that including this part does not alter the results qualitatively). Writing $\Delta_k$ in real space with the non-uniform pairing in a vortex state, and using Eq. S1 for $\Delta_1$, we calculate the local density of states (LDOS) along a line cut approaching the vortex core. We set $t' = 0.0125t$ and $\lambda = 0.025t$. Experimentally, the pocket anisotropy is tiny, and the energy scale of SOC is small, of the order of 2 meV versus $E_f \sim$ 50 meV. Our parameters are in rough correspondence. We set $\xi = 6$, but a comparison to the experimental coherence length would be difficult since the model is not defined on the material lattice. As for the resulting pairing gap $\Delta_0$ (the mean value of the gap on the Fermi surfaces of the two reconstructed bands), we use a relatively larger value to improve the resolution in LDOS, without affecting the qualitative behavior of the bound states, as argued before. In Fig. S9(c) we use $\Delta_0 = 0.2t$. Interestingly, we see a clear ZBCP at the vortex center. Experimentally, the band structure parameters should not change significantly, but the gap size varies slightly from place to place (see Table SII). Nonetheless, the ZBCP is very robust in our STS data. To see whether this is the case in the nodeless $d$-wave scenario, we set $\Delta = 0.26t = 1.3\Delta_0$ -the resulting LDOS is shown in Fig. S9(d) (red curve). Here the conductance peak at the vortex core is shifted away from zero energy. Therefore, the zero mode in Fig. S9(c) is accidental, in contrast to the robustness in the experiment. In fact, we are able to obtain the bound state energy level analytically in the limit of $t' = 0$,

$$E \sim l\omega_0 \pm g, \quad l = 0, \pm 1, \pm 2, \cdots$$

where $g = |g_{k_f}|$ is the energy of SOC on the Fermi surface. The integer angular momentum is a nontrivial result of the spin-momentum texture caused by SOC. Clearly, zero modes may appear but only *accidentally* when g is an integer multiple of $\omega_0$, and such zero modes are not Majorana, since the quasiparticle and its anti-particle have different angular momentum and hence cannot be identical. We remark that the stability of the above energy level requires $g > 0$. If we set $g = 0$ formally, we would have MZMs at $l = 0$ but they are coupled and gapped out by corrections beyond the quasi-classical approximation. This is best seen by ignoring SOC in the first place: The two bands are independent, each manifesting trivial *s*-wave pairing, even though there is a relative negative sign between the pairing in the two bands.

**Other pairing symmetries:** For completeness here we include the vortex states for other pairing symmetries in Fig. S10. Figure S10(a) is for pure *s*-wave in the two-band model as above with the same SOC. There is no ZBCP, similar to the pure *s*-wave case without SOC. Figure S10(b) is for chiral *d*-wave in a simple one-band model. We should mention that on a square lattice the two *d*-wave pairing functions are both one-dimensional irreducible representations of the point group, hence in general they do not mix. We include the chiral *d*-wave here because it is fully gapped and topologically nontrivial. There are gapless edge states in this case. Nonetheless, there is no zero mode in the vortex core. This can actually be understood by solving the vortex problem in the quasi-classical approximation. The result is that the energy levels are similar to that in the pure *s*-wave case for all chiral pairings of even internal angular momentum L ( $L = 0, \pm 2$ for *s*- and chiral *d*-wave), although the bound state wave function depends on. Finally, we present the result for a nodal *d*-wave in Fig. S10(c) for comparison. It is known that in this case the LDOS develops a single broad bump, a kind of resonance, around zero energy and near the vortex core. None of the above cases is consistent with our experimental data.

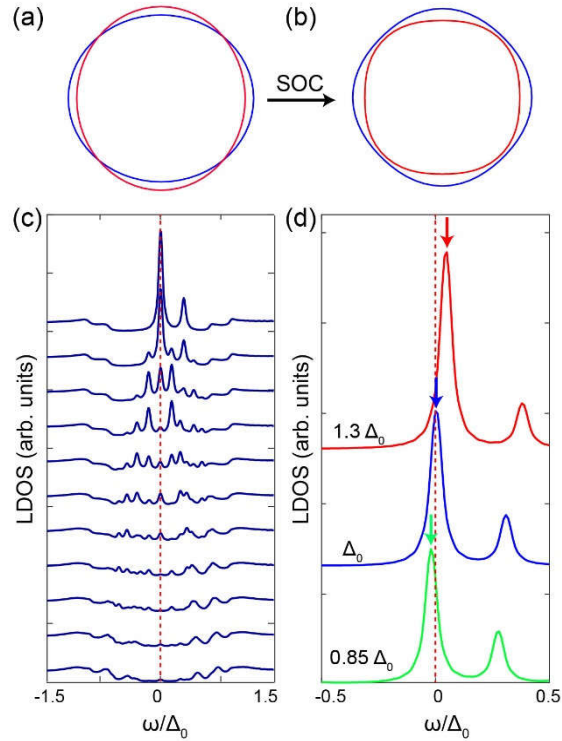

FIG. S9. Model calculations of vortex states for nodeless *d*-wave scenario for $(Li_{1-x}Fe_x)OHFeSe$. (a), (b) Schematic of electron-like Fermi pockets around the *M* point in the Brillouin zone, in the absence (a) and presence (b) of SOC, respectively. The color in (a) shows the nodeless *d*-wave gap function which is positive (red) on one elliptic pocket and negative (blue) on the other. After including SOC, these pockets are reconstructed into the inner (red) and outer (blue) pockets in (b). (c) LDOS along a line-cut approaching (from bottom to top) the vortex center for the bulk gap $\Delta_0$. (d) Comparison of the LDOS with different bulk gaps. The red, blue and green curves respectively represent the calculated LDOS at the vortex core center with a bulk gap of $1.3\Delta_0$, $\Delta_0$ and $0.85\Delta_0$.

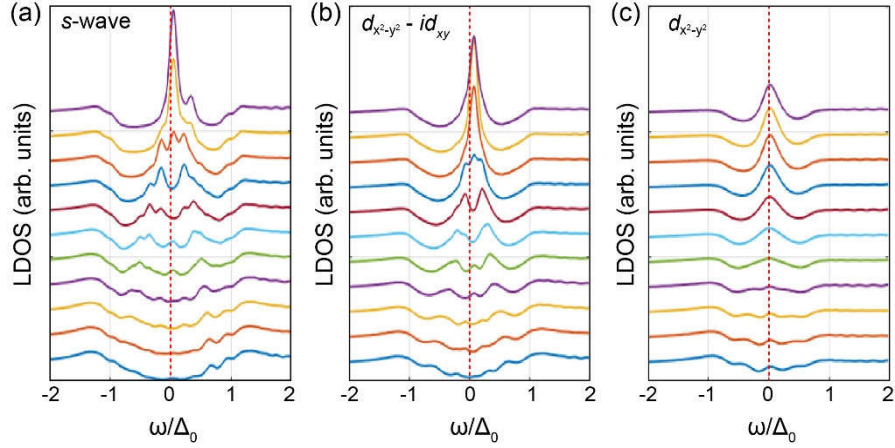

FIG. S10. Model calculations of vortex states for alternative pairing scenarios. LDOS along a line cut approaching the center (from bottom to top) of a positively wound vortex for microscopic pairing symmetries (a) pure *s*-wave with SOC, (b) $d_{x^2-y^2}$-$id_{xy}$-wave and (c) $d_{x^2-y^2}$-wave.

**References:**

S1. A. Georges, G. Kotliar, W. Krauth, and M.J. Rozenberg, *Dynamical mean-field theory of strongly correlated fermion systems and the limit of infinite dimensions.* Rev. Mod. Phys. **68**, 13 (1996).

S2. J. P. Perdew, K. Burke, and M. Ernzerhof, *Generalized Gradient Approximation Made Simple*. Phys. Rev. Lett. **77**, 3865 (1996).

S3. K. Haule, C.-H. Yee, and K. Kim, *Dynamical mean-field theory within the full-potential methods: Electronic structure of $CeIrIn_5$, $CeCoIn_5$, and $CeRhIn_5$.* Phys. Rev. B **81**, 195107 (2010).

S4. Z. P. Yin, K. Haule, and G. Kotliar, *Magnetism and charge dynamics in iron pnictides.* Nature Phys. **7**, 294 (2011).

S5. Z. P. Yin, K. Haule, and G. Kotliar, *Spin dynamics and orbital-antiphase pairing symmetry in iron-based superconductors.* Nature Phys. **10**, 845 (2014).

S6. K. Haule, *Exact Double Counting in Combining the Dynamical Mean Field Theory and the Density Functional Theory.* Phys. Rev. Lett. **115**, 196403 (2015).

S7. P. Werner, A. Comanac, L. de' Medici, M. Troyer, and A. J. Millis, *Continuous-Time Solver for Quantum Impurity Models.* Phys. Rev. Lett. **97**, 076405 (2006).

S8. N. Marzari, A. A. Mostofi, J. R. Yates, I. Souza, and D. Vanderbilt, *Maximally localized Wannier functions: Theory and applications*. Rev. Mod. Phys. **84**, 1419 (2012).

S9. A. A. Mostofi, J. R. Yates, G. Pizzi, Y. Lee, I. Souza, D. Vanderbilt, and N. Marzari, *An updated version of wannier90: A tool for obtaining maximally-localized Wannier functions.* Comput. Phys. Commun.

**185**, 2309 (2014).

S10. R. C. Dynes, V. Narayanamurti, and J. P. Garno, *Direct measurement of quasiparticle-lifetime broadening in a strong-coupled superconductor.* Phys. Rev. Lett. **41**, 1509 (1978).

S11. Z. Y. Du, *et al*. *Sign reversal of the order parameter in $(Li_{1-x}Fe_x)OHFe_{1-y}Zn_ySe$.* Nat. Phys. **14**, 2 (2017).

S12. M. W. Ma, *et al*. *Low-energy spin excitations in $(Li_{0.8}Fe_{0.2})ODFeSe$ superconductor studied with inelastic neutron scattering.* Phys. Rev. B **95**, 100504(R) (2017).

S13. D. F. Agterberg, *et al*. *Resilient nodeless d-wave superconductivity in monolayer FeSe.* Phys. Rev. Lett. **119**, 267001 (2017).